\documentclass[journal=inoraj,manuscript=article]{achemso}
\setkeys{acs}{articletitle = true}
\usepackage[version=3]{mhchem} 
\usepackage{chemformula}
\graphicspath{{Figures/}{\main/Figures/}}

\title{\ch{SiO2}-mediated facile hydrothermal synthesis of spiroffite-type \ch{Co2Te3O8}}
\author{Austin M. Ferrenti}
\affiliation{Department of Chemistry, The Johns Hopkins University, Baltimore, Maryland 21218, USA}
\alsoaffiliation{Institute for Quantum Matter, William H. Miller III Department of Physics and Astronomy, The Johns Hopkins University, Baltimore, Maryland 21218, USA}
\email{*aferren2@jhu.edu}
\author{Natalia Drichko}
\affiliation{Institute for Quantum Matter, William H. Miller III Department of Physics and Astronomy, The Johns Hopkins University, Baltimore, Maryland 21218, USA}
\author{Tyrel M. McQueen}
\affiliation{Department of Chemistry, The Johns Hopkins University, Baltimore, Maryland 21218, USA}
\alsoaffiliation{Institute for Quantum Matter, William H. Miller III Department of Physics and Astronomy, The Johns Hopkins University, Baltimore, Maryland 21218, USA}
\altaffiliation{Department of Materials Science and Engineering, The Johns Hopkins University, Baltimore, Maryland 21218, USA}
\email{*mcqueen@jhu.edu}

\begin{document}
\begin{abstract}
 The hydrothermal synthesis of novel materials typically relies on both knowledge of the redox activities of all cations present in the reaction solution and a small toolset of so-called mineralizers to tune the solution's overall chemical potential. Upon the use of a less conventional mineralizer species, \ch{SiO2}, we show the stabilization of spiroffite-type \ch{Co2Te3O8} under less forceful hydrothermal conditions than in previous reports. When synthesized in the presence of both \ch{SiO2} and each respective alkali carbonate as a secondary mineralizer, silicon substitution in place of tellurium in the host structure becomes apparent, and the corresponding introduced disorder gives rise to enhanced low-temperature ferromagnetism. Our results highlight the complexities of underutilized and combined mineralizer species in the stabilization and tuning of complex magnetic ground states via hydrothermal synthesis techniques.
\end{abstract}

\section*{Introduction}

Hydrothermal synthesis techniques, owing to their typically lower temperature and higher pressure reaction conditions, provide a means of stabilizing solid state structures that would be otherwise too unstable for growth by higher-temperature methods.\cite{mcmillen2016hydrothermal} Materials grown in this way often also possess rarer, more complex lattice geometries, a feature that is of particular interest in the field of frustrated magnetism.\cite{saiduzzaman2021hydrothermal}

The predictable, controlled immobilization of ions into such structures remains an elusive goal of both the geological and materials synthesis communities.\cite{palansooriya2020soil} Although the vast body of known natural mineral phases often provide clues to the means by which synthetic analogues of these systems may be grown, the direct design and synthesis of related materials is generally more complicated.\cite{grohol2003magnetism,ferrenti2023hydrothermal} This typically requires extensive study of the preferential formation and precipitation of different phases under a variety of synthesis conditions, often in the presence of so-called mineralizers.\cite{yang2019conventional} These materials serve to catalyze the formation and growth of stable phases in solution, similar to the natural geologic processes that give rise to mineral formation. While the classes of mineralizers used in laboratory synthesis have been largely restricted to alkali halides and hydroxides, a much greater variety of reactive species exist in natural systems that may also give rise to mineral formation. Over the past several decades, one such material that is ubiquitous in nature, \ch{SiO2}, has been shown to act as a mineralizer in the controlled immobilization of boron and other contaminants in solution.\cite{nozawa2018effect,camenzuli2013immobilization} Dissolved silica has also recently been proposed to be a major mediator in the abiotic formation of the mineral dolomite \ch{CaMg(CO3)2}, a historical process that does not typically occur in the present day.\cite{fang2022dissolved} Given the wide variety of known silicate materials and the suggested importance of silica in non-silicate natural mineral formation, the use of small quantities has the potential to expand the known body of mineralizers for hydrothermal synthesis.

One mineralogically-inspired test case, \ch{Co2Te3O8}, possesses the spiroffite-type structure, consisting of heavily distorted two-dimensional honeycomb layers of \ch{Co^{2+}} cations in an alternating corner- and edge-sharing arrangement that are then bridged by tellurite groups out of plane.\cite{feger1999hydrothermal} The significant distortions of the honeycomb lattice give rise to a complex set of magnetic exchange interactions and antiferromagnetic ordering at \textit{T}$_N$~=~70~K. This material was previously grown from concentrated \ch{NH4Cl} solutions at elevated temperatures ($\sim$~375°C) in hydrothermal autoclaves and more recently via traditional solid state synthesis techniques.\cite{feger1999hydrothermal,li2019structural} In the former study, close attention was paid to the efficacy of various known mineralizer solutions in the nucleation and growth of the \ch{Co2Te3O8}, however only those containing \ch{NH4Cl} were found to result in the desired product.

In this work, we present an alternative, facile means of synthesizing spiroffite-type \ch{Co2Te3O8}, via the inclusion of silica under hydrothermal reaction conditions. While Si content in this composition appears minimal, its optical and magnetic properties vary from those reported for the stoichiometric compound. Upon the further addition of various alkali carbonates to the reaction solution, we observe a gradual tuning of the magnetic ground state from one purely antiferromagnetic in nature to one defined by significantly enhanced ferromagnetic interactions. This strengthening of low-temperature ferromagnetism appears tied to greater defect concentrations in the \ch{Co2Te3O8} structure. This result presents a compelling example of the broad tunability of materials properties that can be achieved through choice of mineralizer.

\section{Materials and methods}
Sample designations denote only the presence (or absence) of each alkali cation in solution during the sample synthesis, and do not imply significant alkali incorporation into the host structure. Sample AF (alkali-free), synthesized with only \ch{SiO2} as the mineralizer, represents the most direct comparison to previously-reported \ch{Co2Te3O8}. 

Polycrystalline \ch{Co2Te3O8} samples were prepared via the hydrothermal reaction of \ch{M^{+}_{2}CO3} [\textbf{Sample AF} = none, \textbf{Sample Li} = \ch{Li2CO3} (ProChem Inc., ACS Grade, $\geq$ 99.0\%), \textbf{Samples Na (1), Na (2)} = \ch{Na2CO3} (Sigma-Aldrich, ReagentPlus\textregistered, $\geq$ 99.5\%), \textbf{Sample K} = \ch{K2CO3}~$\cdot$~{1.5}~\ch{H2O} (ACS Grade, 98.5\%-101.0\%), \textbf{Sample Rb} = \ch{Rb2CO3} (Thermo Scientific, 99\%), \textbf{Sample Cs} = \ch{Cs2CO3} (Alfa Aesar, 99.99\% metals basis)], \ch{CoCO3 $\cdot$ H2O} (Strem Chemicals, 99\%-Co), and \ch{TeO2} (Acros Organics, 99+\%) in a 2~:~2~:~3 molar ratio in the presence of \ch{SiO2}. A total of 300 mg of starting reagents and $\sim$~1.5 mmol (80-90~mg) \ch{SiO2} (Thermo Scientific, 3-12 mm fused lump, 99.99\% metals basis) were added to a 23-mL Teflon-lined hydrothermal reaction vessel (Parr Instrument Company, Model \#4749) with 12 mL of deionized water. The solution was then sparged with Argon gas for 6 minutes to reduce the dissolved \ch{O2} concentration and the vessel was quickly sealed. The vessel was placed into a furnace preheated to 200°C, allowed to dwell for 72 hours, and finally air-quenched to room temperature. Once cooled, the product solution was vacuum-filtered and rinsed with additional deionized water before drying in air. Sample characterization was performed on as-synthesized samples that were dried at 200°C in air for at least 6 hours to remove surface moisture before being quenched in a dessicator, unless otherwise specified. 

The majority of samples possessed small white spots in the otherwise bright purple powder corresponding to a minor \ch{TeO2} impurity phase. Sample AF was also slightly darker purple in color than Samples Li~-~Cs. The post-reaction mass of the input \ch{SiO2} lump was observed to have decreased by up to 5\% in the case of Sample AF and up to 17\% across all other samples. Where unspecified, data shown for Sample "Na" is representative of both Sample Na (1) and Sample Na (2). Both samples were prepared four months apart from one another, under nominally identical conditions, with the same precursor materials, and in the same reaction vessel and drying oven. This, as well as the observed similarities between the properties of Samples Na (1) and Li and Samples Na (2) and K~-~Cs, respectively, suggests that this variability is due to factors that otherwise tune the overall chemical potential of the solution, such as small changes in the pH of the deionized water or minor additional hydration of precursor materials over time. Where different, data for Sample Na (2) is presented in the Supplementary Information for clarity. 

Polycrystalline \ch{Zn2Te3O8} samples were prepared via the hydrothermal reaction of \newline \ch{Zn5(CO3)2(OH)6} (Alfa Aesar, 97\%) and \ch{TeO2} (Acros Organics, 99+\%) in a 2~:~3 molar ratio in the presence of $\sim$~1.5 mmol (85 mg) \ch{SiO2} ((Thermo Scientific, 3-12 mm fused lump, 99.99\% metals basis). A total of 300 mg of starting reagents were added to a 23-mL Teflon-lined hydrothermal reaction vessel (Parr Instrument Company, Model \#4749) with 12 mL of deionized water and $\sim$~1 mL of 0.3 M citric acid solution before sparging with Argon for 6 minutes and heating under identical conditions to those used to synthesize the Co-containing samples. Samples were found to consist largely of the target \ch{Zn2Te3O8} phase, with small $\sim$~3.5 wt\% \ch{ZnTeO3} and $\sim$~0.5 wt\% \ch{TeO2} impurities, estimated via Rietveld refinement.

Powder X-ray diffraction (pXRD) patterns were collected on a laboratory Bruker D8 Focus diffractometer with LynxEye detector and Cu K$\alpha$ radiation in the 2$\theta$ range from 5-60°. Rietveld refinements on sample pXRD data with an internal silicon standard (Si powder, Acros Chem, 99.9\%) were performed using Topas 5.0 (Bruker). Structures were visualized using Vesta 3.\cite{momma2011vesta} The morphology and chemical composition of the as-synthesized samples were probed using the energy dispersive x-ray spectroscopy (EDS) option of a JEOL JSM-IT100 scanning electron microscope (SEM), with a 20 kV acceleration voltage. Simultaneous thermogravimetric analysis/differential thermal analysis (TGA/DTA) for each sample was performed using a TA Instruments Q600 SDT. The samples were loaded into pre-dried alumina pans and heated to 900°C at a rate of 10°C/minute under inert \ch{N2} flow (100 mL/minute). Furnace heating was then stopped and the samples cooled naturally under \ch{N2} gas flowing at the same rate.

Fourier-transform infrared (FTIR) spectra of representative \ch{Co2Te3O8} powder samples were collected using a Thermo Scientific Nicolet iS5 FTIR with an iD5 ATR attachment from 500 to 4000~cm$^{-1}$. Raman scattering spectra were measured in the backscattering geometry over the range from 100-4500~cm$^{-1}$ using a Horiba JY T64000 spectrometer equipped with an Olympus microscope, using a 2 micron diameter, $\sim$1 mW laser probe. The spectra were excited using the 514.5 nm line of a Coherent Innova 70C Spectrum laser. Representative spectra were collected from powder samples heated for at least 6 hours at 200°C and calibrated with the primary optical phonon mode (at 520~cm$^{-1}$) of a silicon standard to ensure a reliable comparison between samples. No signs of laser-induced degradation of the samples were observed over the course of each measurement. 

Magnetization data was collected on a Quantum Design Magnetic Property Measurement System (MPMS3). Magnetic susceptibility was approximated as magnetization divided by the applied magnetic field ($\chi\approx M/H$). Curie-Weiss fittings of the measured magnetic susceptibility were performed assuming the mass of \ch{Co2Te3O8} remains constant for all samples. 

Heat capacity measurements were performed using a Quantum Design Physical Property Measurement System (PPMS) by the semi-adiabatic pulse technique with a 1\% temperature rise and measurement over three time constants. Heat capacity samples were prepared by cold pressing a thin pellet of each sample at 2 tons of applied pressure, which was then cut to fit the dimensions of the heat capacity platform. The change in magnetic entropy as a function of temperature, $\Delta S_{\mathrm{mag}}$, was approximated as $\Delta S_{\mathrm{mag}}$~=~$\int C_{\mathrm{p}}/T\,dT$ of the measured $C_{\mathrm{p,mag}}$ for each sample ($C_{\mathrm{p, total, \ch{Co2Te3O8}}}$ - $C_{\mathrm{p, total, \ch{Zn2Te3O8}}}$), from \emph{T}~=~0--65 K. The entropy rise from \emph{T}~=~0~K to 2~K was estimated from linear extrapolation over this range. To account for the mass difference between Zn and Co and thus more accurately substract the phononic contribution to the measured heat capacity of the \ch{Co2Te3O8} sample over the relevant temperature range, the temperature scale of the \ch{Zn2Te3O8} sample was scaled by $\frac{\Theta^3_{\ch{Co2Te3O8}}}{\Theta^3_{\ch{Zn2Te3O8}}}$ = ($\frac{\text{Molar mass of \ch{Zn2Te3O8}}}{\text{Molar mass of \ch{Co2Te3O8}}}$)$^{3/2}$ = ($\frac{\text{628.7 g/mol}}{\text{641.6 g/mol}}$)$^{3/2}$ = 0.97. An additional 12\% scaling of the measured \ch{Zn2Te3O8} heat capacity was found to be necessary to ensure convergence with the Co-analogues above \textit{T}~=~65~K.

\section{Results and Discussion}
\subsection*{Structural and compositional characterization}
\begin{figure}
  \hspace*{-0.5cm}
  \includegraphics[width=1.0\textwidth]
  {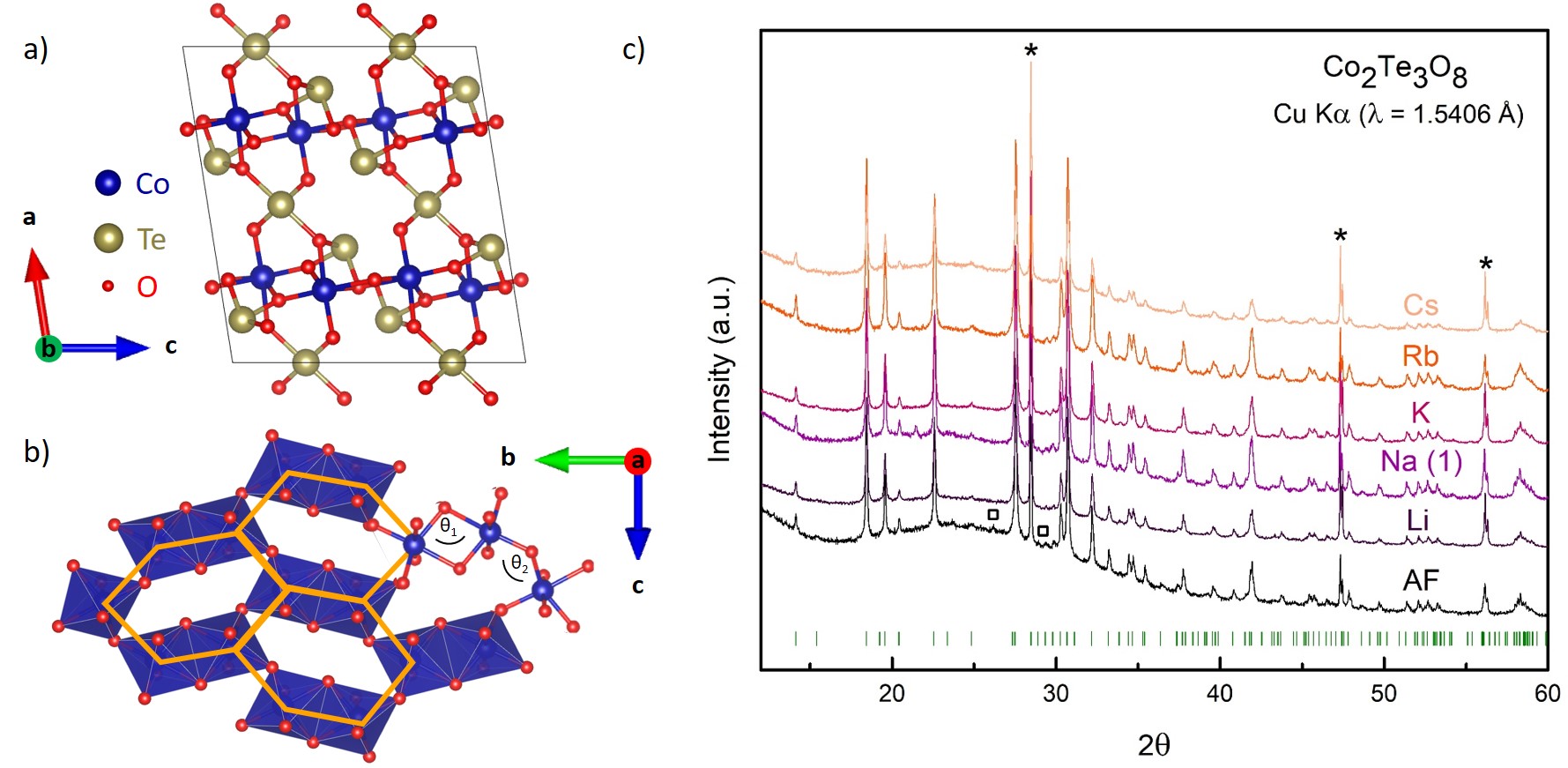}
  \caption{\textbf{} Crystal structure of \ch{Co2Te3O8} a) perpendicular to and b) within the honeycomb (\textit{bc}) plane. The black line denotes the unit cell and the orange hexagons highlight the highly distorted cobalt honeycomb. $\theta_1$ and $\theta_2$ denote the Co-O4-Co and Co-O3-Co bond angles mediating magnetic superexchange. c) pXRD patterns of representative \ch{Co2Te3O8} powder samples synthesized in the presence of \ch{SiO2} and \ch{M^{+}_{2}CO3} [M = none (black), Li (dark purple), Na (1) (magenta), K (red), Rb (orange), Cs (salmon)] - hereafter referred to as Samples AF, Li, Na, K, Rb, and Cs. Sample designations denote only the presence (or absence) of each alkali cation in solution during the sample synthesis, and do not imply significant alkali incorporation into the host structure. All patterns agree reasonably well with that of the spiroffite-type \ch{Co2Te3O8} phase, indicating only subtle changes to the known structure when synthesized by the methods described in this work. The black stars and squares correspond to the internal silicon standard and small \ch{TeO2} impurity phase, respectively.}
  \label{All_pXRD}
\end{figure}

Although previously synthesized by traditional solid state and high-pressure hydrothermal synthesis methods, the addition of a small quantity of \ch{SiO2} to the reaction vessel was found to stabilize the monoclinic \ch{Co2Te3O8} structure under more facile hydrothermal conditions. The powder diffraction pattern of Sample AF, synthesized in this manner and shown in black in Figure \ref{All_pXRD}c, agrees well with the known monoclinic structure (space group $C2/c$ [15]), with a small decrease in the refined lattice parameters and only minor changes to the bond angles mediating magnetic exchange (Figure S1, Tables S1-S4). However, the broad, amorphous hump visible below 2$\theta$~=~40° suggests either the presence of an amorphous impurity phase or a more significantly defected structure.

More significant average changes are observed when \ch{Co2Te3O8} samples are grown in the presence of both \ch{SiO2} and alkali carbonate. For Samples Li-Cs, the refined lattice parameters and Co-O3-Co and Co-O4-Co bond angles are also slightly reduced, relative to pure \ch{Co2Te3O8} (Tables S1, S4, Figures S2-S7). With the possible exception of \ch{Li^+}, significant alkali incorporation into the structure's open channels would be unlikely, suggesting that the presence of alkali cations instead serves to modulate the chemical potential of the solution and introduce greater disorder. Interestingly, despite the more significant average structural changes suggested by the Rietveld refinements of the diffraction data for Samples Li-Cs, only Samples Na (1), Rb, and Cs display a similar broad, amorphous hump to that observed for Sample AF.

To determine whether the compositions synthesized via the method reported in this work deviate from the known \ch{Co2Te3O8} stoichiometry suggested by the pXRD data, energy dispersive x-ray (EDS) point and map spectra were collected for all samples. Point spectra collected from several distinct regions of Sample AF suggest the presence of two distinct phases - one that agrees well with the expected 2~:~3 Co~:~Te ratio and shows no sign of the characteristic silicon K$\alpha$ peak at 1.74 keV, and another that appears to possess minimal Te and a variable Co~:~Si ratio (Figure S9, Table S5). As there is no sign of any known cobalt silicate phase in the pXRD data for Sample AF (Figure S1), these EDS point spectra likely correspond to the presence of a small, amorphous cobalt-doped \ch{SiO2} phase. This phase is also observed in Sample Li, alongside small, variable quantities of Si incorporated into the \ch{Co2Te3O8} structure, with the calculated atomic ratios suggesting that it substitutes for Te as \ch{Co2Te$_{3-x}$Si$_x$O8} (Figure S10, Table S5). Similarly, the EDS point spectra recorded for Sample Na (1) support a distribution of Si levels in \ch{Co2Te$_{3-x}$Si$_x$O8}, where x ranges from 0.12-1.44 and the overall Co~:~(Te+Si) ratio remains around 2~:~3 (Figure S11, Table S5). By contrast, the point spectra measured from Samples Na (2) - Cs largely suggest the presence of one brighter, lower-substituted \ch{Co2Te$_{3-x}$Si$_x$O8} phase, where x $\leq$ 0.1, with a small quantity of darker polycrystalline lumps with larger x (Figures S12-S15, Tables S5, S6). The Si distributions observed for each sample then split the compositions into three distinct categories - Sample AF which appears to show little to no Si incorporation into the \ch{Co2Te3O8} structure, Samples Li and Na (1), which contain \ch{Co2Te3O8} with significant Si substitution for Te, and Samples Na (2) - Cs, which largely consist of lesser-substituted \ch{Co2Te$_{3-x}$Si$_x$O8}, with a few Si-rich clusters. 

This is particularly evident in the EDS map spectra taken across each sample, where Samples A.F and Na (2) - Cs show only minimal (0.3-3.0~atomic~\%) average Si content across the region analyzed, whereas this increases to 6.5\% and 7.2\% for Samples Li and Na (1), respectively (Figures S16-S22, Table S7). While the Co and Te distributions appear relatively uniform throughout each sample, the Si content appears clustered in smaller regions, further supporting the presence of several distinct \ch{Co2Te$_{3-x}$Si$_x$O8} phases with variable x. In some regions, such as that shown in Figure S23 for Sample Rb, small chunks of \ch{SiO2} are also clearly distinguished from the polycrystalline \ch{Co2Te$_{3-x}$Si$_x$O8}, albeit without the significant Co content suggested from the various point spectra. 

Due to these small compositional differences, minor changes in the thermodynamics of step-wise dehydration and higher-temperature decomposition for each sample are also apparent in the thermal analysis (Figures S24-S25). On heating, Sample AF exhibits a slow, consistent mass decrease up to 800°C, before the onset of greater mass loss that coincides with a minor exotherm in the measured DTA curve, indicative of melting and/or sample decomposition. All other samples exhibit several distinct mass drops on heating over the same range, as well as a shifting of the DTA exotherm temperature. However, the general behavior remains consistent, suggesting that all samples undergo at least one dehydration step, followed by eventual incongruent melting. These mass decreases correspond to a loss of at least 1.6 molar equivalents of \ch{H2O} from the structure in the case of Sample AF to at most 2.7 molar equivalents for Samples Na (2) and Rb.

\begin{figure}
  \includegraphics[width=1.0\textwidth]
  {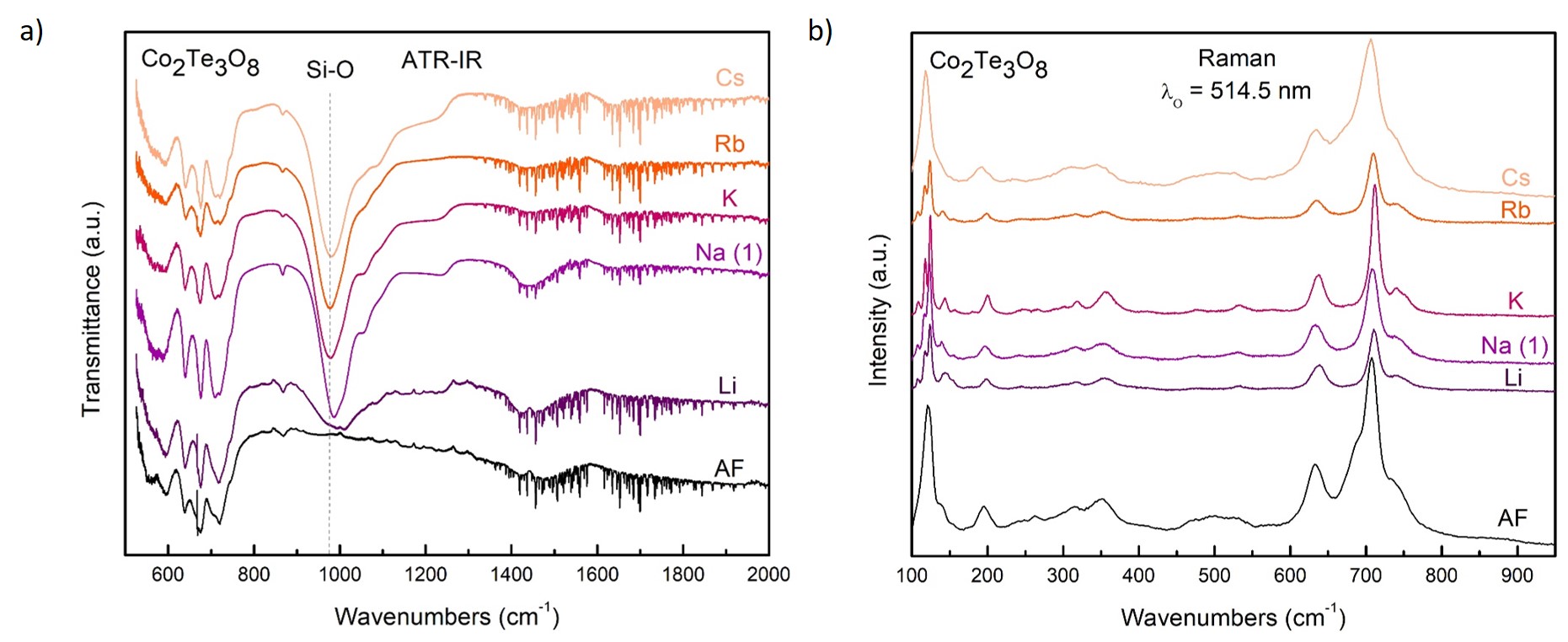}
  \caption{\textbf{} a) Infrared and b) Raman scattering spectra of Samples AF (black), Li (dark purple), Na (1) (magenta), K (red), Rb (orange), and Cs (salmon). With the exception of Sample AF (alkali-free), all infrared spectra agree well with that reported for natural spiroffite (\ch{Mn2Te3O8}), with slight shifts in band frequencies due to the slight difference in mass between Mn and Co. The dashed black line indicates the frequency of the typical Si-O stretch. The measured Raman spectra for all samples also agree well with that reported for natural spiroffite, with the expected frequency shifts for the synthetic Co-analogue. The bands in the Sample AF and Sample Cs spectra are significantly broadened, relative to the others, indicating more significant disorder.}
  \label{IR_Raman}
\end{figure}

Despite the significant silicon substitution suggested by the EDS spectra, the measured infrared and Raman scattering spectra agree well with those reported for natural spiroffite, \ch{(Mn,Zn)2Te3O8}, accounting for the mass difference between Co and Mn, as well as for synthetic \ch{Co2Te3O8} (Figure \ref{IR_Raman}).\cite{feger1999hydrothermal} Between 600-1300~cm$^{-1}$, all infrared bands in natural spiroffite were previously attributed to Te-O vibrational modes, with no mention of features at higher frequencies. For \ch{Co2Te3O8} samples synthesized in the presence of \ch{SiO2}, these lower-frequency bands are indeed present and largely in agreement between compositions, with the exception of the weak band centered around 718~cm$^{-1}$. For Samples Na - Cs, two weak humps appear at 707 and 720~cm$^{-1}$, while for Sample Li only the latter is visible and for Sample AF the second weak hump is pushed to 700~cm$^{-1}$. As Samples Li and Na (1) likely possess the highest Si content and Sample AF the least, these band splittings likely correspond to more random disorder around the site of the corresponding Te-O vibration. 

The most pronounced difference between the infrared spectra of each sample can be seen in the band centered around 975~cm$^{-1}$ for Samples K - Cs, which is observed to harden to 980~cm$^{-1}$ for Sample Na, weaken and broaden for Sample Li, and be entirely absent in the case of Sample AF. This vibration has often been observed in natural samples of spiroffite and several related tellurite phases, generally being attributed to \ch{TeO3}$^{2-}$ stretches.\cite{frost2009raman,frost2010application} However, this also coincides with the range of common Si-O vibrational frequencies in similar systems.\cite{he1995vibrational} These features are consistent with the ubiquity of silicate intergrowths in natural mineral samples, as well as with the absent Si content observed in Sample AF. If this were simply a result of a lowering of the local symmetry of one of the crystallographically-distinct Te sites, this band would likely be split and exhibit frequency shifts, but instead it is absent for Sample AF. We thus attribute this band to the Si-O stretch of Si introduced into the \ch{Co2Te3O8} structure. Relative to as-synthesized Sample Li, the intensity of this measured vibrational mode also appears to decrease on prolonged air exposure or heating to 200°C, however this is likely due to the sampling of distinct regions with varying Si-content during each measurement (Figure S28b).

The small differences in band splittings apparent between the infrared spectra of each sample become less pronounced overall in the measured Raman spectra, with the exception of those for Samples AF and Cs. The frequencies and relative intensities of all vibrational modes for Samples Li - Rb are consistent with those reported at nearly ambient-pressure for \ch{Co2Te3O8} synthesized via traditional solid state synthesis methods.\cite{li2019structural} The spectra for Samples AF and Cs exhibit significantly more broadening and possess several additional vibrational modes, indicative of a more disordered structure.\cite{pena2020structural}. However, this broadening of peaks in both the measured Raman spectra and pXRD patterns is most apparent for Samples AF and Cs, both of which possess the lowest silicon content. Given the significant difference between the ionic radii of 4-fold coordinate Te$^{4+}$ and Si$^{4+}$, much more significant lattice distortions and shrinkage of the unit cell would be expected, particularly for the more Si-rich phases. The introduced, much smaller Si should result in a more highly disordered structure than is observed here. The synthesis of this phase at lower hydrothermal pressures than in previous reports suggests that introduced Si likely serves to relieve internal steric pressure, allowing for the stabilization of the \ch{Co2Te3O8} structure under less extreme conditions and only a more significantly disordered system in its absence. \cite{feger1999hydrothermal}

The structural and compositional differences observed between samples can likely be attributed to the relative ionic activities of dissolved silica and the respective alkali carbonates in aqueous solution. As the solubility of \ch{SiO2} and subsequently the number of Si$^{4+}$ cations entering the solution are both quite low, its sole presence in the case of Sample AF would be expected to result in the lowest incorporated Si content and least significant structural distortions and changes, relative to the other compositions. However, this is not observed. While the solubilities of \ch{Li2CO3} and \ch{Na2CO3} in water are both higher than for \ch{SiO2} and lower than those of the larger alkali carbonates, the reaction solution should become more basic overall, which apparently enables the introduction of small amounts of Si into the host structure in place of Te. For Samples K - Cs, the relatively high solubilities of the respective carbonates in \ch{H2O} produce even more basic solutions, albeit with lesser Si incorporation. Typically, the solubility of \ch{SiO2} in \ch{H2O} increases as a function of pH, with the formation of \ch{H4SiO4} and later dissociation to \ch{H^+} and \ch{H3SiO4^-}.\cite{palmer1995geochemistry} However, given that Si solubility and \ch{TeO2} co-precipitation with the title phase was generally low for all samples, the structural changes observed across this series most likely arise from a gradual tuning of the chemical potential of the reaction solution. This small adjustment encourages increased substitution of \ch{Si^{4+}} cations for \ch{Te^{4+}} under hydrothermal reaction conditions, but is insufficient to promote higher-energy oxidation or reduction of these cations, relative to the precursor materials.

\subsection*{Magnetic properties}
The structural and compositional changes observed across this series also give rise to significant deviations from the magnetic response of the known \ch{Co2Te3O8} phase. Spiroffite-type \ch{Co2Te3O8} synthesized under higher pressure hydrothermal conditions orders antiferromagnetically at \textit{T}$_N$~=~70 K, with a Curie-Weiss temperature of $\theta_{\text{CW}}$~=~-112.0~K, suggesting strong antiferromagnetic exchange. However, this net antiferromagnetism arises from competing exchange interactions between the alternating corner- and edge-sharing Co$^{2+}$ octahedra that comprise the heavily distorted honeycomb lattice. The Co-O-Co bond angles defining both superexchange pathways deviate significantly from the ideal 90° and 180° cases expected to result in purely ferromagnetic or antiferromagnetic interactions, although they lie closer to the former case. The presence of additional complex exchange pathways throughout the structure further complicates determination of the known material's magnetic structure, however the average structural changes observed for samples synthesized in the presence of \ch{SiO2} provide some clues.

\begin{figure}
  \includegraphics[width=1.0\textwidth]
  {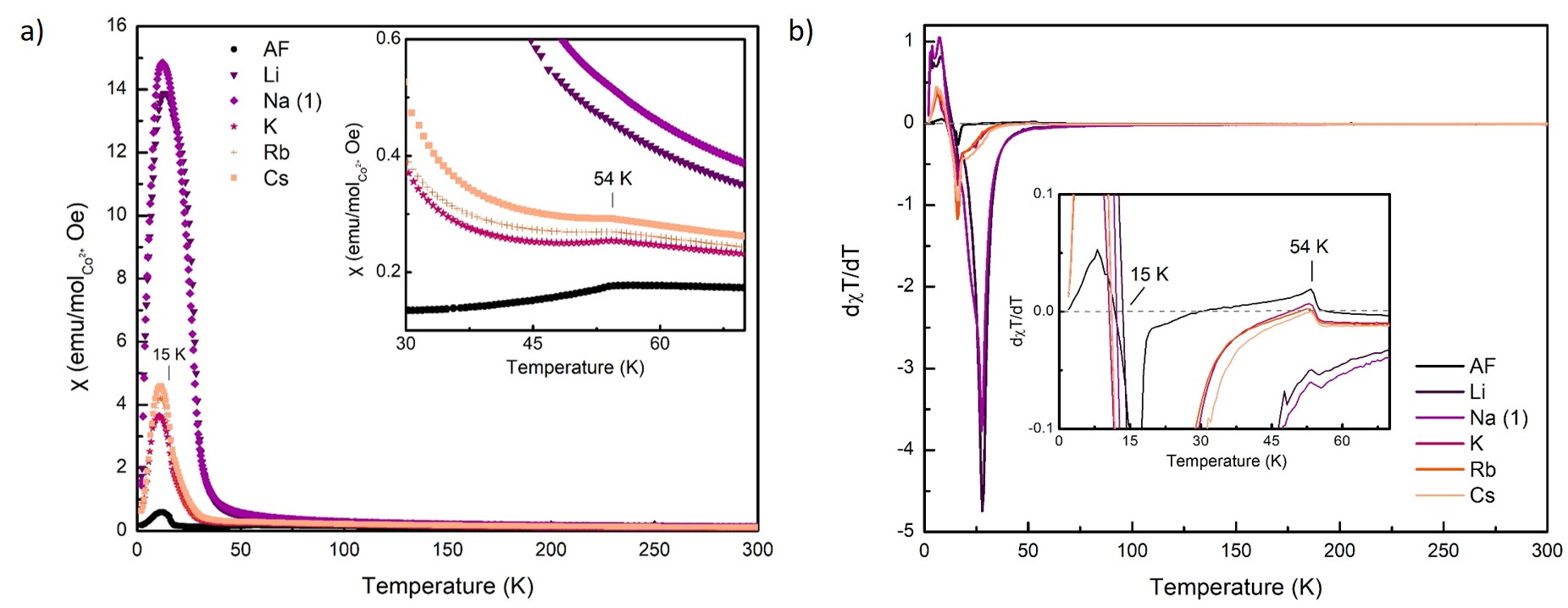}
  \caption{\textbf{} a) Temperature dependent magnetic susceptibility and b) its derivative times temperature with respect to temperature of Samples AF (black), Li (dark purple), Na (1) (magenta), K (red), Rb (orange), Cs (salmon). All samples deviate from the known, purely antiferromagnetic \ch{Co2Te3O8} phase ($T_\text{N}$~=~70~K) synthesized via a high-pressure hydrothermal method, and instead exhibit signs of increased competition between antiferromagnetic and ferromagnetic exchange in the structure. Sample Na (1) exhibits a similar response to Sample Li, while Sample Na (2) (Figure S29) behaves more akin to Samples K, Rb, and Cs.} 
  \label{Mag_comb}
\end{figure}

Sample AF, which possesses the most similar average structure to that of pure \ch{Co2Te3O8} and minimal \ch{SiO2} incorporation, undergoes two magnetic transitions as a function of temperature, as seen in Figure \ref{Mag_comb}a and derived from the d$\chi$T/dT vs. T plot in Figure \ref{Mag_comb}b. The first, occurring at \textit{T}~=~53.4~K, resembles the AFM ordering transition of previously-synthesized \ch{Co2Te3O8}, while the other occurs around \textit{T}~=~16.2~K, is significantly stronger, and appears more ferromagnetic (FM) in nature. The presence of the AFM transition at a lower temperature than in the known material, as well as the appearance of the new FM transition, suggest either a more balanced distribution of net FM and AFM exchange interactions or greater disorder-induced spin canting, despite the minimal average structural changes derived from the pXRD data. The calculated effective magnetic moment for this composition, $p_{\text{eff}}$~=~4.07(1)~$\mu_\text{B}$, is also 20\% lower than that reported for pure \ch{Co2Te3O8} and analogous to the expected spin-only moment for a high-spin Co$^{2+}$ system, 3.88~$\mu_\text{B}$ (Table \ref{CW_fits_all_dried}). Interestingly, despite the appearance of the ferromagnetic-like transition in the measured DC susceptibility, the calculated Curie-Weiss temperature for this composition decreases to $\theta_{\text{CW}}$~=~-150.5(4)~K, indicative of a strengthening of AFM correlations.

Samples Li-Cs, synthesized in the presence of both \ch{SiO2} and alkali carbonate, behave quite similarly to one another and deviate more significantly from the behavior expected for a pure AFM. Relative to Sample AF, Samples Na (2), K, Rb, and Cs display an enhanced FM transition at \textit{T}~=~16.2 K, while the AFM kink at \textit{T}~=~53.4 K remains static, suggesting a further tuning of the magnetic exchange pathways to favor stronger net FM exchange, albeit still in competition with strong AFM interactions (Figure \ref{Mag_comb}a). This is also supported by the average changes observed in the bond angles comprising the major superexchange pathways. The average narrowing by 1.9° and 6.6°, with respect to the parent phase, is expected to decrease the degree of overlap between the Co 3d and O 2p orbitals mediating superexchange (Table S4). AFM superexchange along both pathways would thus be expected to weaken, while FM superexchange would be strengthened. The calculated effective magnetic moments for these samples are similar to that observed for Sample AF, ranging from $p_{\text{eff}}$~=~3.89(1)-4.37(1) ~$\mu_\text{B}$, while the calculated Curie-Weiss temperature for each is either similar or reduced relative to that of pure \ch{Co2Te3O8} (Table \ref{CW_fits_all_dried}). This is again consistent with a net, minor weakening of AFM exchange.

The measured magnetic susceptibility for Samples Li and Na (1) deviates most radically from the behavior reported for \ch{Co2Te3O8}, with each displaying a sequence of magnetic transitions and magnetization enhanced by over an order of magnitude (Figures \ref{Mag_comb}a, \ref{Mag_comb}b). While the average structural changes and calculated effective magnetic moments are similar to those observed for Samples Na (2)~-~Cs, the calculated Curie-Weiss temperature drops to $\theta_{\text{CW}}$~=~-~42.1(2)~K~and~-~44.3(2)~K for Samples Li and Na (1), respectively (Figures \ref{Mag_comb}a, \ref{Mag_comb}b; Table \ref{CW_fits_all_dried}). Despite the Curie-Weiss temperatures remaining negative, indicating the preservation of dominant AFM interactions, FM exchange in these samples is clearly significantly enhanced. Additionally, all samples display a distinct splitting between the magnetic susceptibility curves for \textit{T}~$\leq$~20~K when measured after cooling in the absence (ZFC) or presence (FC) of an applied magnetic field (Figure S29). This is indicative of spin freezing that sets in only with the dominance of net ferromagnetic interactions between neighboring Co$^{2+}$ moments and is quite common among disordered magnetic systems.\cite{zheng2008defect}

\begin{figure}
  \includegraphics[width=1.0\textwidth]
  {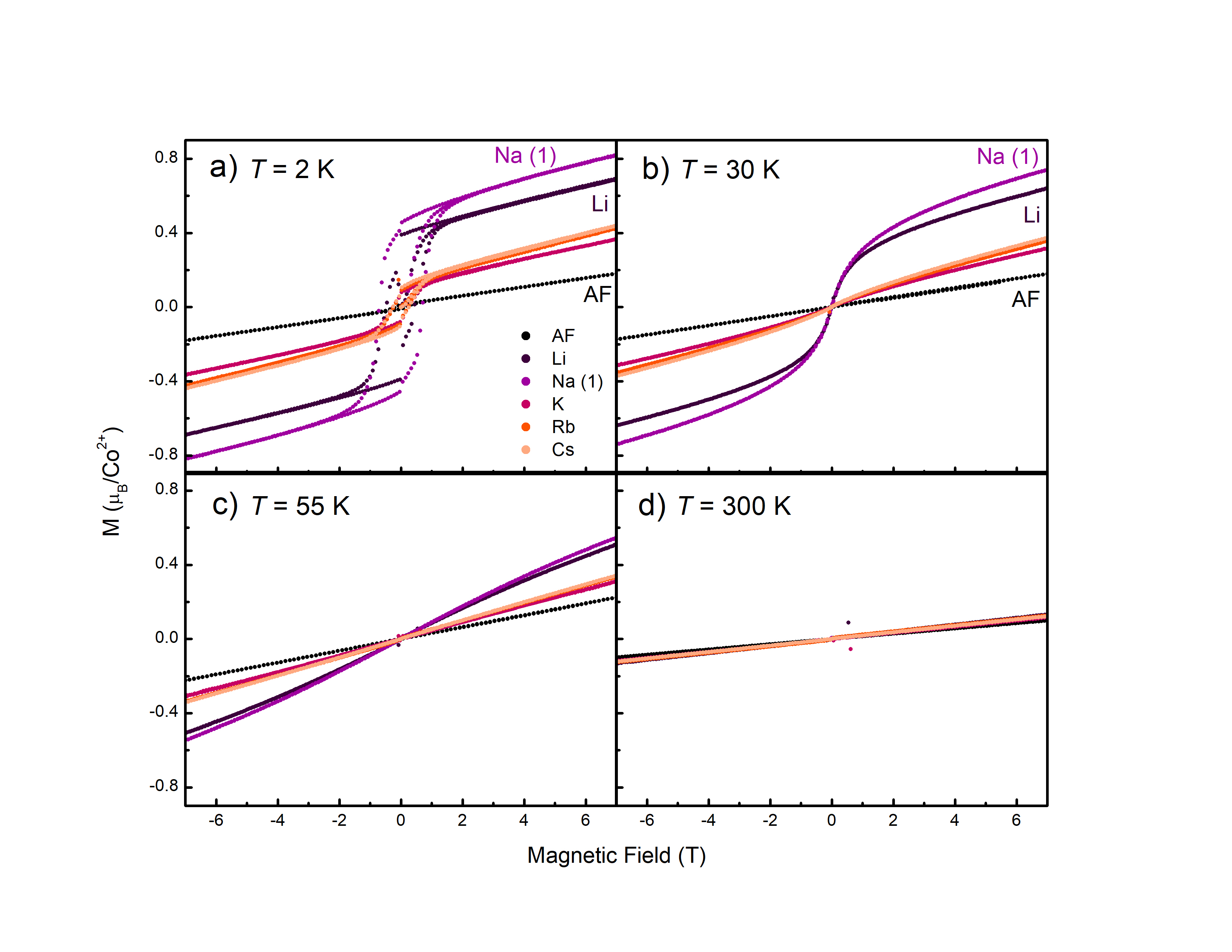}
  \caption{\textbf{}Field-dependent magnetization at a) $T$~=~2~K, b) $T$~=~30~K, c) $T$~=~55~K, and d) $T$~=~300~K of Samples AF (black), Li (dark purple), Na (1) (magenta), K (red), Rb (orange), and Cs (salmon). Minor hysteresis is observed for all but the alkali-free sample at $T$~=~2~K, with the Li and Na (1) samples exhibiting the largest coercivity and magnetization.} 
  \label{Magnetization_comb}
\end{figure}

The observed enhancement of ferromagnetic interactions alongside changing solution potential is also apparent in the field-dependent magnetization of each sample at \textit{T}~=~2~K, as seen in Figure \ref{Magnetization_comb}a. The magnetization of Sample AF exhibits a largely linear field-dependence, with only minor coercivity up to $\pm$~0.5~T. Similar to the trends observed in the temperature-dependent magnetic susceptibility, Samples K-Cs and Na (2) (Figure S30b) exhibit a greater magnetization and coercivity out to $\pm$~1.0~T, while these features are further amplified for Samples Li and Na (1). At \textit{T}~=~30~K (Figure \ref{Magnetization_comb}b), Sample AF exhibits a purely linear response, while all other samples exhibit a superparamagnetic-like curvature as a function of increasing applied field strength, the magnitude of which is similar to that measured at \textit{T}~=~2~K (Figure \ref{Magnetization_comb}a). A similar superparamagnetic response has been commonly observed in hydrothermally-synthesized cobalt-based nanoparticle systems, however much less frequently in samples where the typical particle size is on at least the micrometer scale, as is the case for the samples reported in this work.\cite{rath2011magnetic,fayazzadeh2020magnetic} Such behavior was recently also observed in a boron imidazolate-based metal-organic framework, where this curvature was attributed to a disordered collection of FM an AFM exchange.\cite{davis2024tunable} Just above the AFM ordering transition, nearly all samples display only a linear response at \textit{T}~=~55~K, as well as in the fully paramagnetic state at \textit{T}~=~300~K (Figure \ref{Magnetization_comb}), with Sample Na (1) showing a slight curvature. Up to the maximum applied field strength of $\mu_0$H~=~$\pm$~7~T, none of the compositions are observed to saturate, reaching at most only $\sim$~16~\% of the effective magnet moment in the case of Sample Na (1).

In order to estimate the impact of heat treatment on the magnetic response of \ch{Co2Te3O8} samples, Sample Na (1) was also measured both prior to the 200°C heating and after additional annealing at 600°C. As-synthesized Sample Na (1) possesses a weaker net magnetization and lower-temperature onset of the ZFC-FC splitting (Figure S30a), relative to the same sample after heating to 200°C, as well as a $\sim$~25\% reduction in both the calculated Curie-Weiss temperature, $\theta_{\text{CW}}$~=~--32.3(3)~K, and effective magnetic moment, $p_{\text{eff}}$~=~3.00(1)~$\mu_\text{B}$ (Table S8). Similar behavior was observed for as-synthesized Samples K and Rb (Table S8). After heating to 600°C, the measured magnetic susceptibility further decreases in magnitude and the ZFC-FC splitting is pushed to higher temperatures, indicative of a more disordered system (Figure S30a). The calculated effective magnetic moment, $p_{\text{eff}}$~=~3.48(1)~$\mu_\text{B}$, and Curie-Weiss temperature, $\theta_{\text{CW}}$~=~-39.3(3)~K, lie in between those calculated for the as-synthesized and 200°C-heat samples. The field-dependent magnetization of both the as-synthesized and 600°C-heat samples agrees well with that observed for the 200°C-heat sample, with similar coercivities (Figure S30b).

\subsection*{Specific heat}
\begin{figure}
  \includegraphics[width=1.0\textwidth]
  {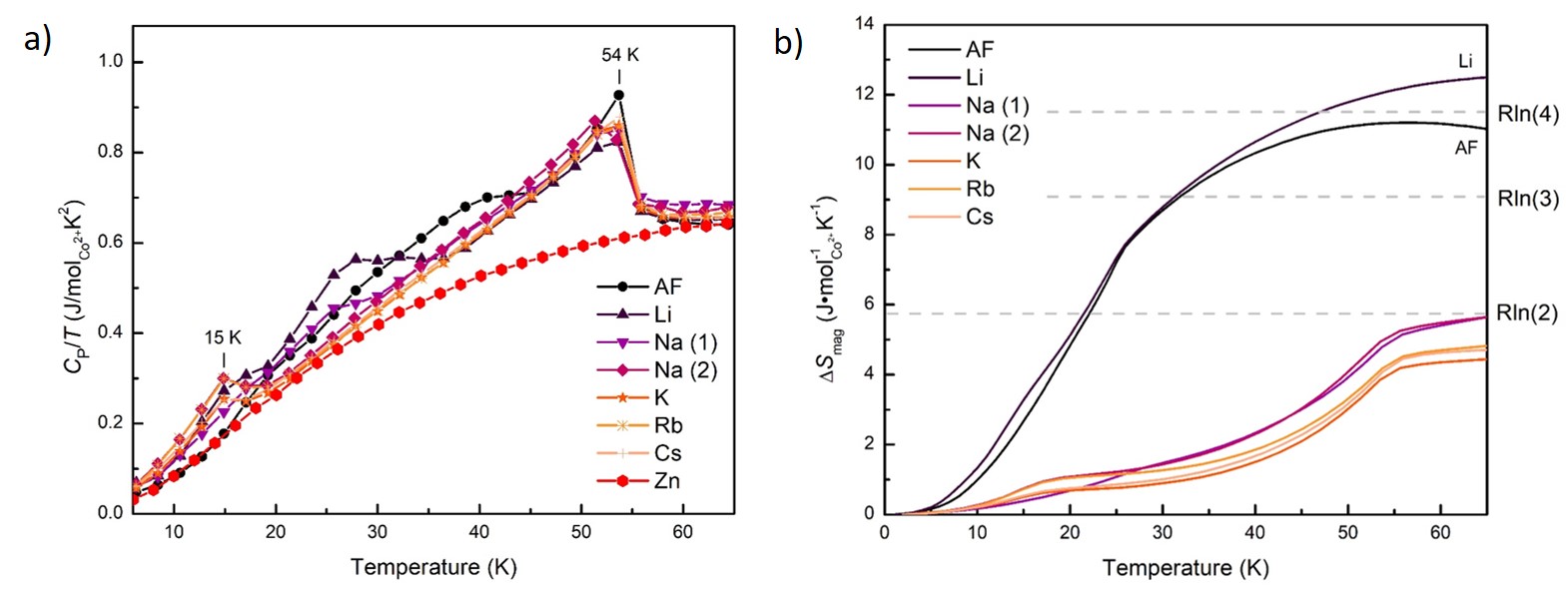}
  \caption{\textbf{} a) Heat capacity divided by temperature as a function of temperature of Samples AF (black), Li (dark purple), Na (1) (magenta), Na (2) (pink), K (orange), Rb (tangerine), Cs (salmon), and \ch{Zn2Te3O8} (red). The observed humps correspond well to the transitions in the measured DC magnetic susceptibility. b) Magnetic entropy rise as a function of temperature of Samples AF (black), Li (dark purple), Na (1) (magenta), Na (2) (pink), K (orange), Rb (tangerine), Cs (salmon). Samples AF and Li saturate at around the Rln(4) limit expected for a high-spin Co$^{2+}$, \textit{S}~=~3/2 system, while Samples Na (1)-Cs saturate closer to the Rln(2) expected for a low-spin Co$^{2+}$, \textit{S}~=~1/2 system.} 
  \label{CpT_deltaS_comp}
\end{figure}

The discrepancies observed between the magnetic response of each \ch{Co2Te3O8} sample are also reflected in the measured specific heat, shown in Figure \ref{CpT_deltaS_comp}. All samples display a relatively sharp thermal anomaly centered at around \textit{T}~=~53.7~K, in agreement with the resilience of the relatively weak AFM ordering transition apparent from magnetic susceptibility measurements. Samples Li, K, Rb, and Cs also exhibit a second peak centered around \textit{T}~=~15.0~K, corresponding to the FM ordering feature observed. Interestingly, Samples AF, Li, and Na (1) exhibit broader features between these two peaks, suggesting a more diffuse release of magnetic entropy and subsequently more disordered magnetism on this temperature scale. This division between the two sample sets is especially evident when comparing the measured specific heats of Samples Na (1) and Na (2) (Figure S30c). Whereas Sample Na (2) exhibits the two sharper peaks corresponding to those observed in the magnetic susceptibility data, Sample Na (1) lacks the lower-temperature feature, having broadened and been pushed to higher temperatures (Figure S30a).

The phononic contribution to the measured heat capacity was approximated as the scaled heat capacity of the non-magnetic \ch{Zn2Te3O8} analogue and subtracted from that measured for each Co-containing sample, $C_{\mathrm{Total}}$, to estimate the magnetic contribution, $C_{\mathrm{mag}}$, to the total heat capacity. In order to ensure good agreement between the two curves above \textit{T}~=~60 K, the measured heat capacity of the nonmagnetic analogue was additionally scaled by 12\%. Integration of $C_{\mathrm{mag}}$ from \emph{T}~$\sim$~0-65~K for Samples AF and Li yields a rise in magnetic entropy of $\Delta S_{\mathrm{mag}}$~=~12.50~$J\cdot mol_{Co^{2+}}^{-1}K^{-1}$ and $\Delta S_{\mathrm{mag}}$~=~11.21~$J\cdot mol_{Co^{2+}}^{-1}K^{-1}$, respectively, in good agreement with that expected for high-spin \ch{Co^{2+}}, where $\Delta S_{\mathrm{mag}}$ = Rln(2S+1) = 11.5 $J\cdot mol^{-1}K^{-1}$ for a \emph{S}~=~3/2 system (Figure \ref{CpT_deltaS_comp}b). For Samples Na (1)-Cs, the magnetic entropy begins to saturate between $\Delta S_{\mathrm{mag}}$~=~4.43~$J\cdot mol_{Co^{2+}}^{-1}K^{-1}$ and $\Delta S_{\mathrm{mag}}$~=~5.62~$J\cdot mol_{Co^{2+}}^{-1}K^{-1}$, closer to the expected low-spin \ch{Co^{2+}} value ($\Delta S_{\mathrm{mag}}$ = Rln(2S+1) = 5.8 $J\cdot mol^{-1}K^{-1}$) for a \emph{S}~=~1/2 system (Figure \ref{CpT_deltaS_comp}b). The choice of solution potential then allows for the tuning of \ch{Co2Te3O8} from a high-spin to low-spin \ch{Co^{2+}}  state. The disagreement observed between the magnetic response and magnetic entropy rise of some samples, particularly in the case of Samples Na (1) and Na (2), can likely be attributed to variability in the degree of distortion of the local Co$^{2+}$ coordination environment, giving rise to a larger net magnetization for Sample Na (1) than for Sample Na (2).

\section*{Conclusions}
In conclusion, we present the study of a series of spiroffite-type \ch{Co2Te3O8} compositions synthesized hydrothermally in the presence of \ch{SiO2} and various alkali carbonates. As the choice of mineralizer and overall synthetic pathway is adjusted across the series, a gradual tuning of the dominant magnetic exchange pathways is observed, manifesting in the suppression of the known antiferromagnetic ordering temperature and emergence of weak low-temperature ferromagetism. Compositional analysis suggests that these changes arise from variable substitution of Si for Te in the host \ch{Co2Te3O8} structure, giving rise to a range of disordered analogues that are synthesizable under less aggressive hydrothermal reaction conditions than has been previously reported. Interestingly, the measured pXRD, infrared and Raman spectra for the majority of compositions suggest that silicon substitution for tellurium actually serves to relieve internal steric pressure and stabilize the \ch{Co2Te3O8} structure, while disorder is much greater at lower substitution levels. Overall, this series of compositions represents a compelling example of the means by which novel and modified phases may be more controllably precipitated from hydrothermal solution to tune materials properties.

\bibliography{Refs.bib}
\clearpage

\newpage

\newpage

\newpage

\begin{table}
    \caption{\textbf{Curie-Weiss fitting parameters obtained from analysis of \ch{Co2Te3O8} powder magnetization measurements from \emph{T}~=~70-300~K.}}
    \label{CW_fits_all_dried}
    \begin{tabular}{|c|c|c|c|}
    \hline
    \textbf{Sample} & \textbf{$\theta_{\text{CW}}$} & \textbf{$p_{\text{eff}}$} & \textbf{$\chi_0$} \\ 
    & (K) & ($\mu_\text{B}$/\ch{Co^{2+}}) & \\\hline
    Literature \ch{Co2Te3O8} & -112.0 & 5.13 & - \\\hline
    AF & -150.5(4) & 4.07(1) & -2.2(1)~x~10$^{-3}$ \\\hline
    Li & -42.1(2) & 3.82(1) & -3.1(3.1)~x~10$^{-5}$ \\\hline
    Na (1) & -44.3(2) & 3.99(1) & -3.1(3.1)~x~10$^{-5}$ \\\hline
    Na (2) & -105.8(3) & 4.01(1) & -7.6(3)~x~10$^{-4}$ \\\hline
    K & -93.2(3) & 3.89(1) & -3.1(3.1)~x~10$^{-5}$ \\\hline
    Rb & -91.5(3) & 3.96(1) & -3.1(3.1)~x~10$^{-5}$ \\\hline
    Cs & -118.2(3) & 4.37(1) & -3.1(3.1)~x~10$^{-5}$ \\ \hline
    \end{tabular}
\end{table}
\newpage

\newpage

\newpage

\section{Acknowledgments}
This work was supported by the Institute for Quantum Matter, an Energy Frontier Research Center funded by the U.S. Department of Energy, Office of Science, Office of Basic Energy Sciences, under Grant DE-SC0019331. The MPMS3 system used for magnetic characterization was funded by the National Science Foundation, Division of Materials Research, Major Research Instrumentation Program, under Award \#1828490. The authors would like to thank T. Soetojo and J. Davis for technical assistance. 
\textbf{Data and materials availability:} All data needed to evaluate the conclusions in the paper are present in the paper and/or the Supplementary Materials. All data underlying this study will be made openly available at the online repository 10.34863/nc3n-3c33.

\clearpage
\end{document}


\newpage
\tableofcontents
\newpage
\section{pXRD refinement data}

\begin{figure}
  \includegraphics[width=1.0\textwidth]
  {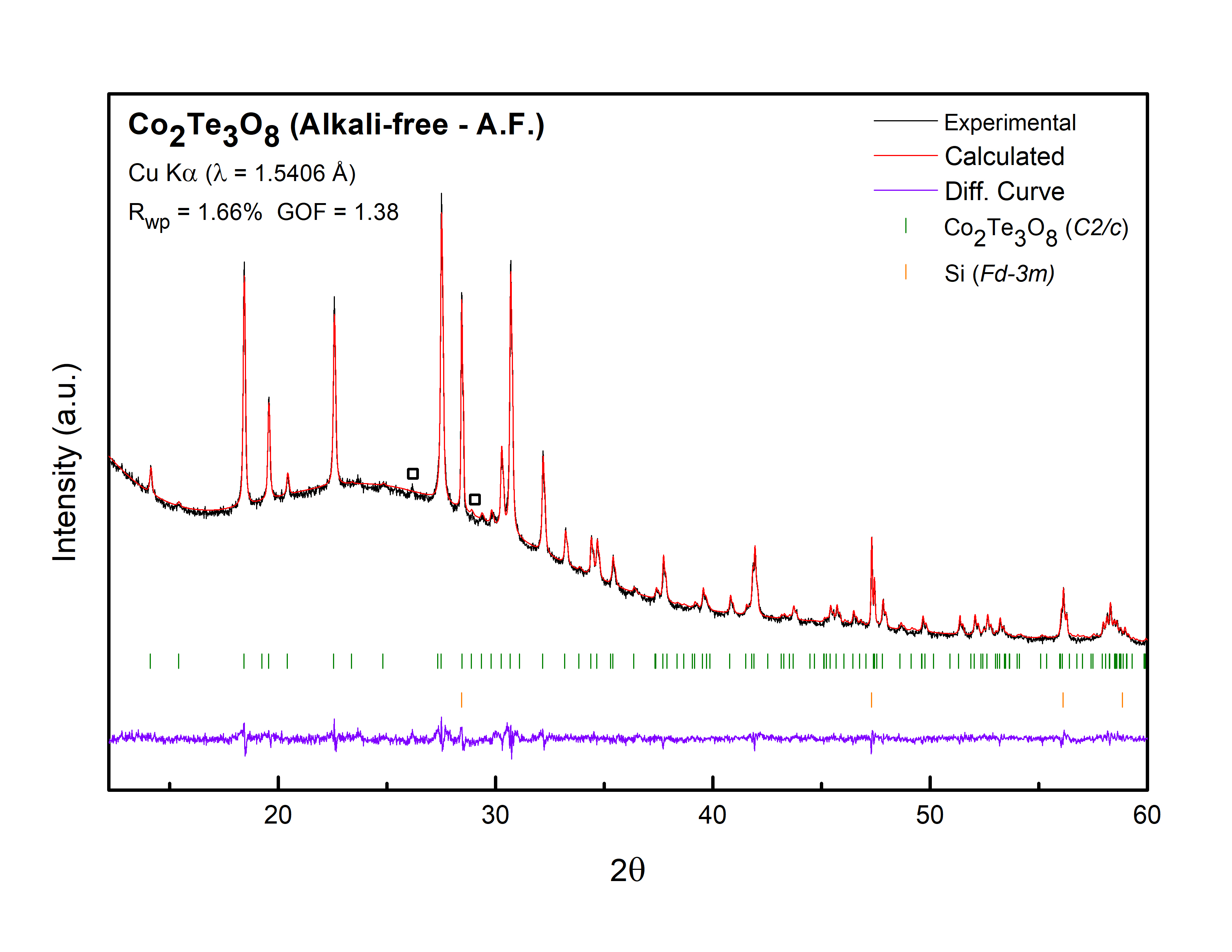}
  \caption{\textbf{} pXRD pattern of a representative \ch{Co2Te3O8} sample synthesized in the presence of \ch{SiO2} (black). Two-phase simulated Rietveld refinement profile for \ch{Co2Te3O8} ($C2/c$) and \ch{Si} ($Fd\Bar{3}m$) (red), and difference curve between the experimental and simulated patterns (purple). The black squares denote a small \ch{TeO2} ($P4_12_12$) impurity phase.} 
  \label{pXRD_AF}
\end{figure} 

\begin{figure}
  \includegraphics[width=1.0\textwidth]
  {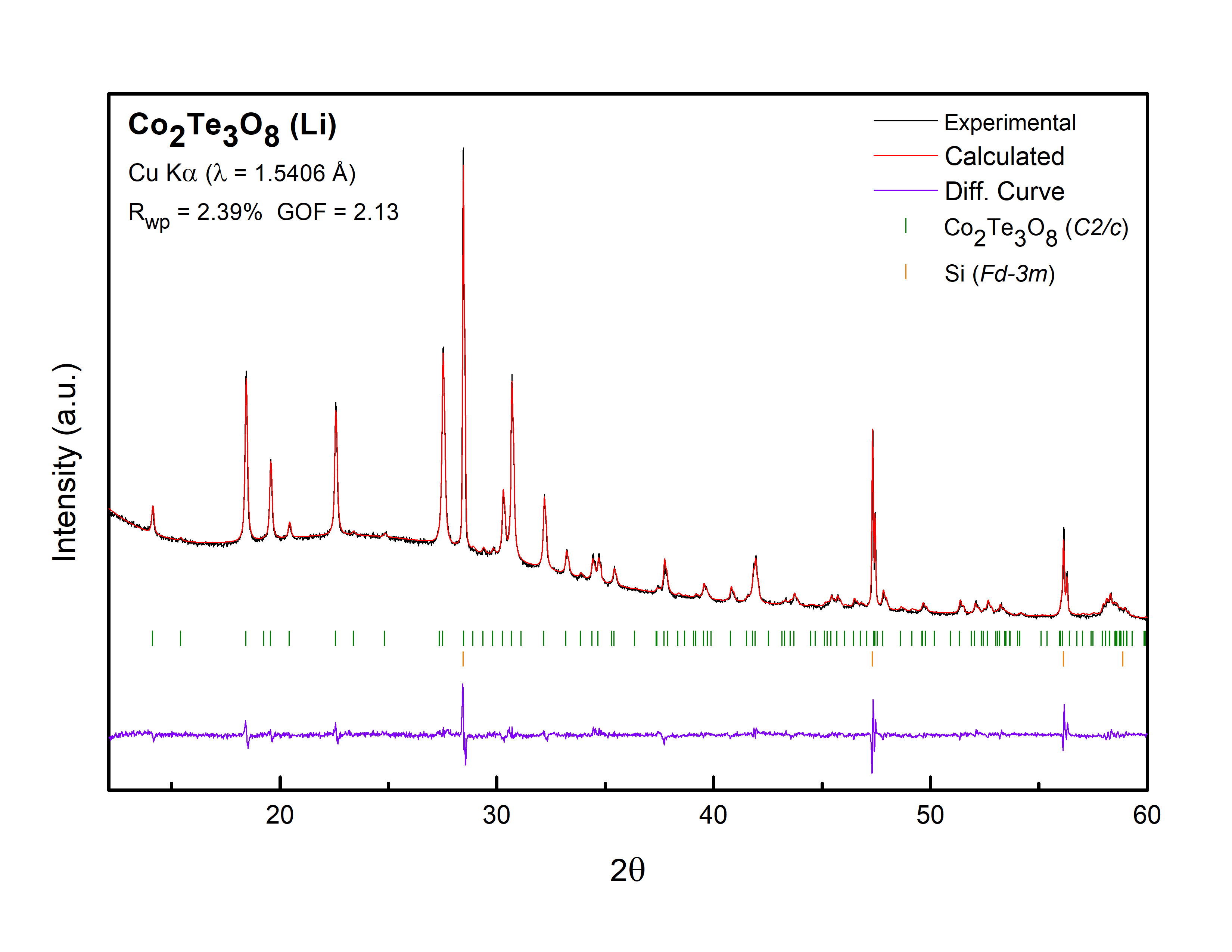}
  \caption{\textbf{} pXRD pattern of a representative \ch{Co2Te3O8} sample synthesized in the presence of \ch{SiO2} and \ch{Li2CO3} (black). Two-phase simulated Rietveld refinement profile for \ch{Co2Te3O8} ($C2/c$) and \ch{Si} ($Fd\Bar{3}m$) (red), and difference curve between the experimental and simulated patterns (purple).} 
  \label{pXRD_Li}
\end{figure} 

\begin{figure}
  \includegraphics[width=1.0\textwidth]
  {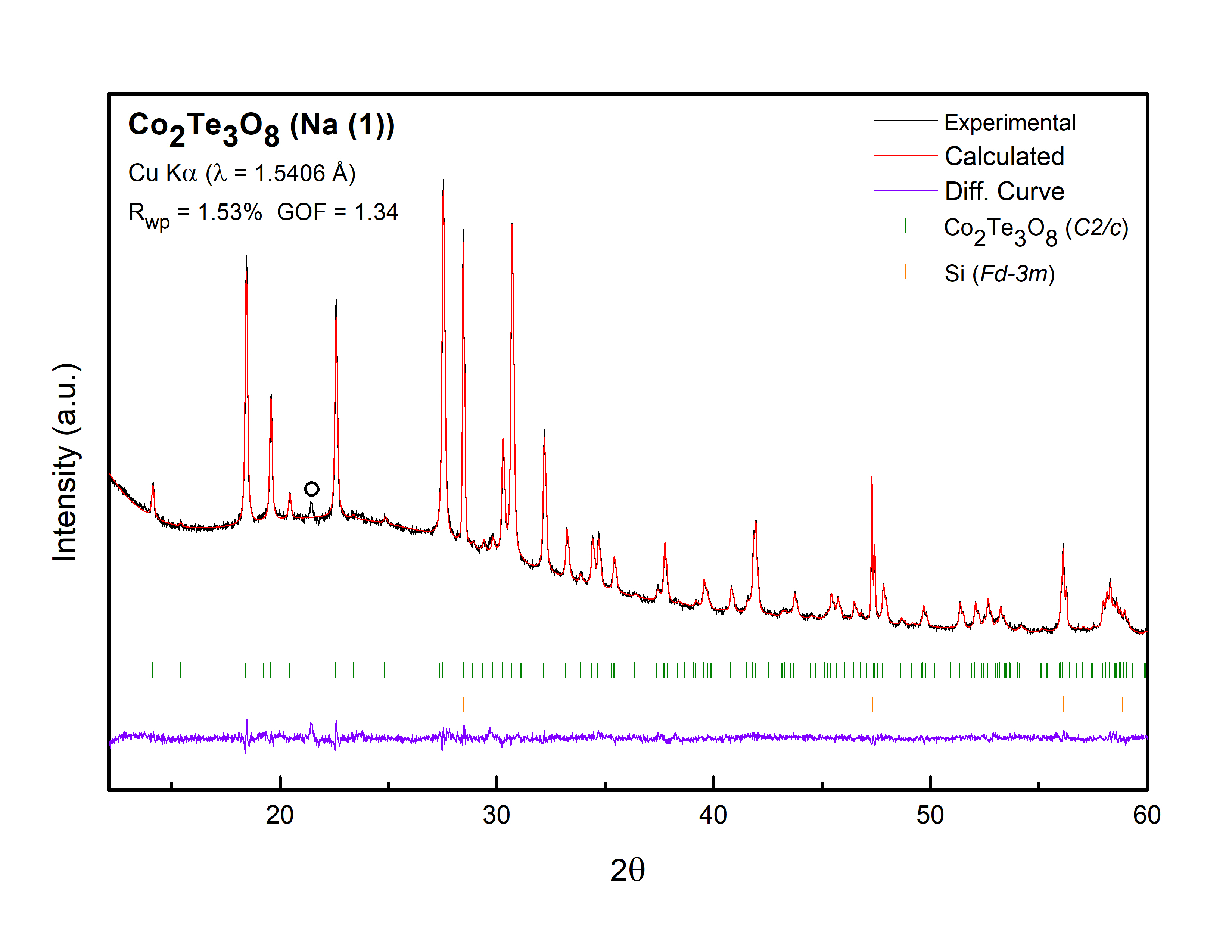}
  \caption{\textbf{} pXRD pattern of one representative \ch{Co2Te3O8} sample synthesized in the presence of \ch{SiO2} and \ch{Na2CO3} (black). Two-phase simulated Rietveld refinement profile for \ch{Co2Te3O8} ($C2/c$) and \ch{Si} ($Fd\Bar{3}m$) (red), and difference curve between the experimental and simulated patterns (purple). The black circle denotes an unknown impurity phase.} 
  \label{pXRD_Na_1}
\end{figure} 

\begin{figure}
  \includegraphics[width=1.0\textwidth]
  {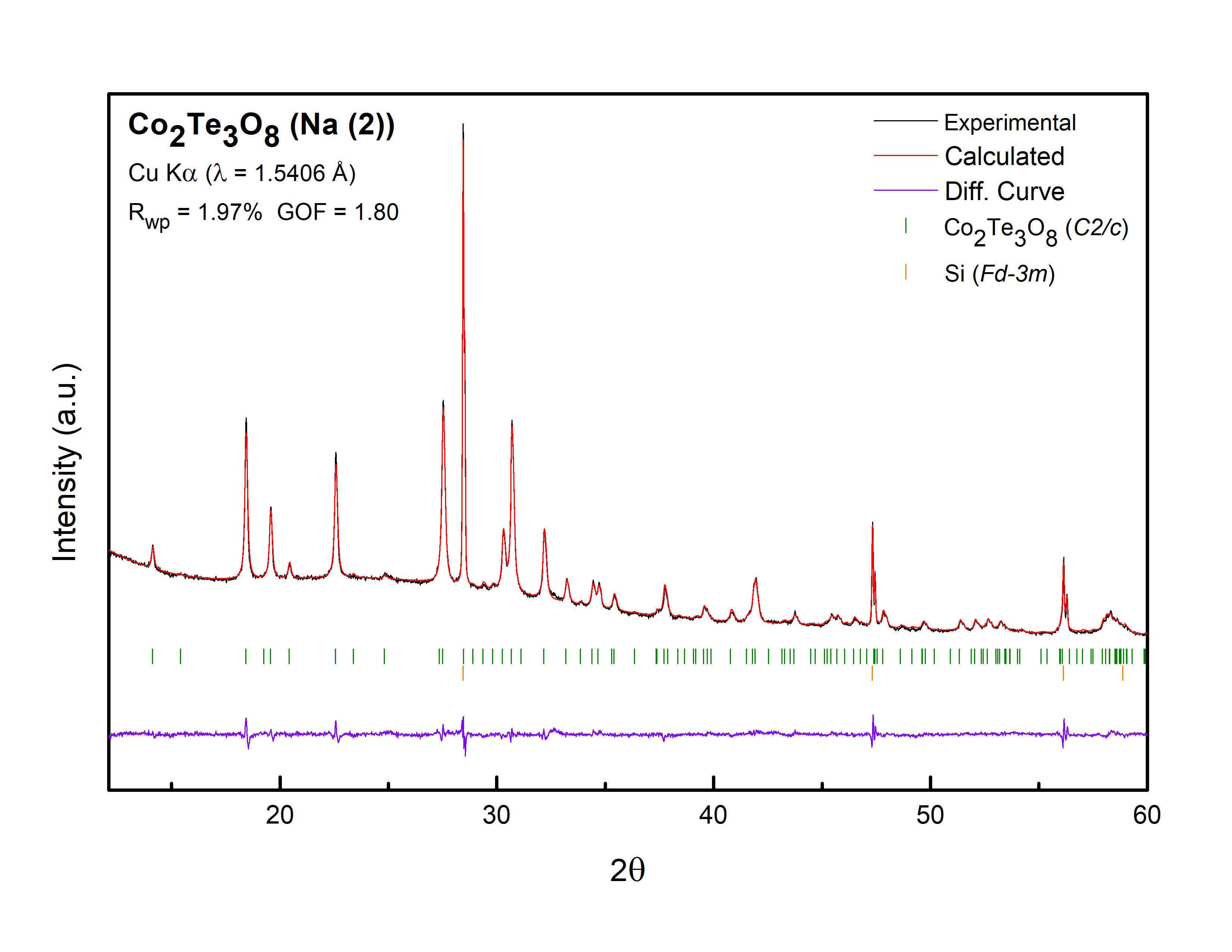}
  \caption{\textbf{} pXRD pattern of a second representative \ch{Co2Te3O8} sample synthesized in the presence of \ch{SiO2} and \ch{Na2CO3} (black). Two-phase simulated Rietveld refinement profile for \ch{Co2Te3O8} ($C2/c$) and \ch{Si} ($Fd\Bar{3}m$) (red), and difference curve between the experimental and simulated patterns (purple).} 
  \label{pXRD_Na_2}
\end{figure} 

\begin{figure}
  \includegraphics[width=1.0\textwidth]
  {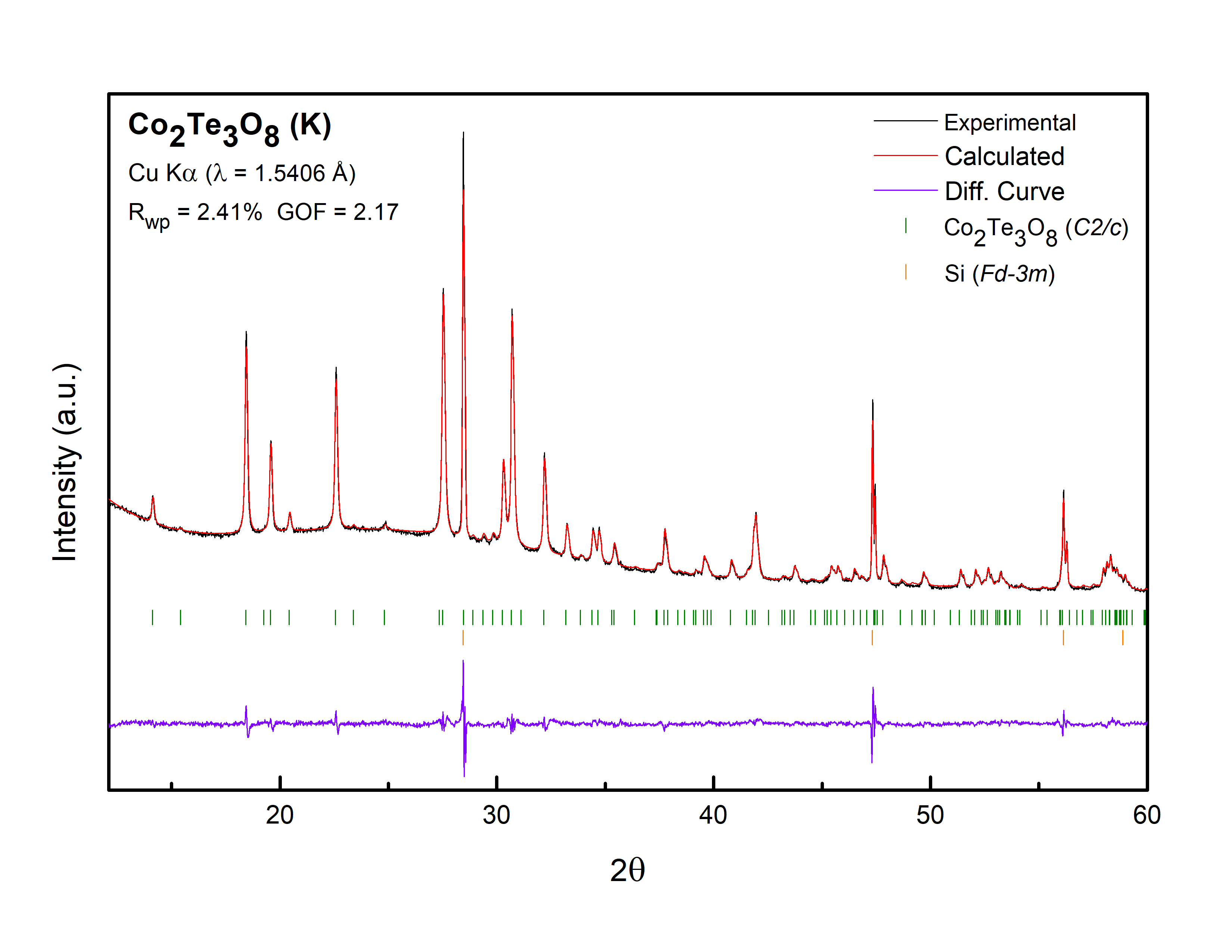}
  \caption{\textbf{} pXRD pattern of a representative \ch{Co2Te3O8} sample synthesized in the presence of \ch{SiO2} and \ch{K2CO3} (black). Two-phase simulated Rietveld refinement profile for \ch{Co2Te3O8} ($C2/c$) and \ch{Si} ($Fd\Bar{3}m$) (red), and difference curve between the experimental and simulated patterns (purple).} 
  \label{pXRD_K}
\end{figure}

\begin{figure}
  \includegraphics[width=1.0\textwidth]
  {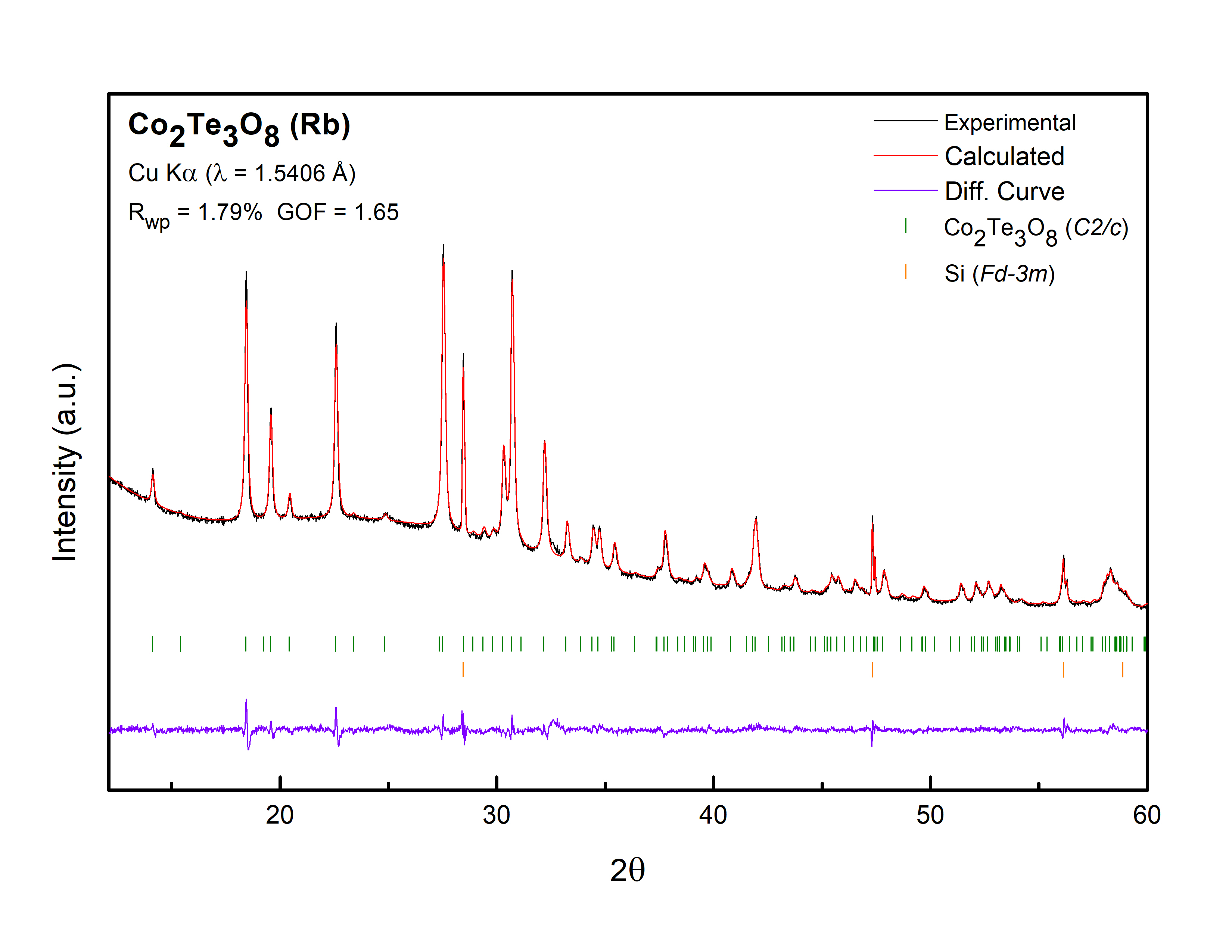}
  \caption{\textbf{} pXRD pattern of a representative \ch{Co2Te3O8} sample synthesized in the presence of \ch{SiO2} and \ch{Rb2CO3} (black). Two-phase simulated Rietveld refinement profile for \ch{Co2Te3O8} ($C2/c$) and \ch{Si} ($Fd\Bar{3}m$) (red), and difference curve between the experimental and simulated patterns (purple).} 
  \label{pXRD_Rb}
\end{figure}

\begin{figure}
  \includegraphics[width=1.0\textwidth]
  {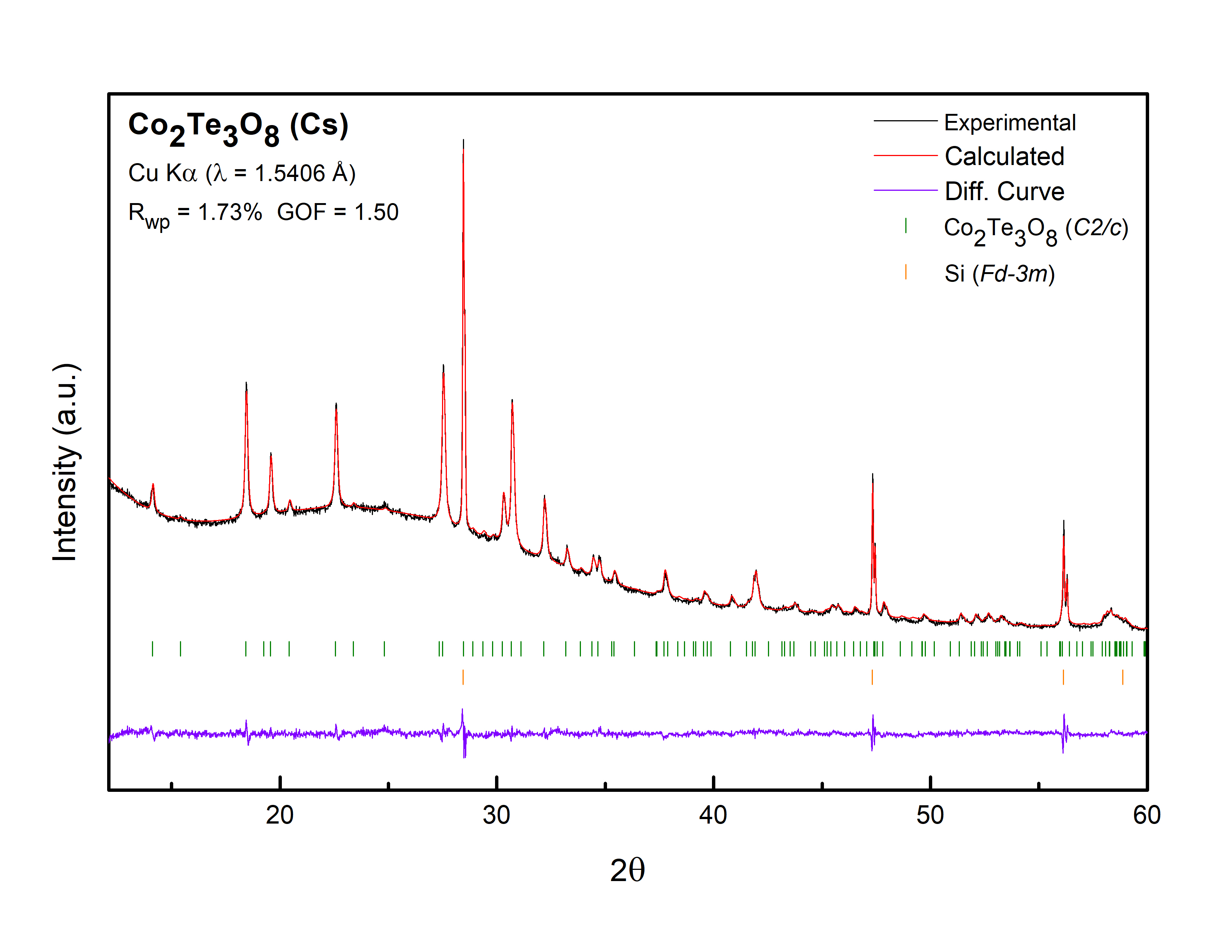}
  \caption{\textbf{} pXRD pattern of a representative \ch{Co2Te3O8} sample synthesized in the presence of \ch{SiO2} and \ch{Cs2CO3} (black). Two-phase simulated Rietveld refinement profile for \ch{Co2Te3O8} ($C2/c$) and \ch{Si} ($Fd\Bar{3}m$) (red), and difference curve between the experimental and simulated patterns (purple).} 
  \label{pXRD_Cs}
\end{figure} 
\clearpage

\begin{figure}
  \includegraphics[width=1.0\textwidth]
  {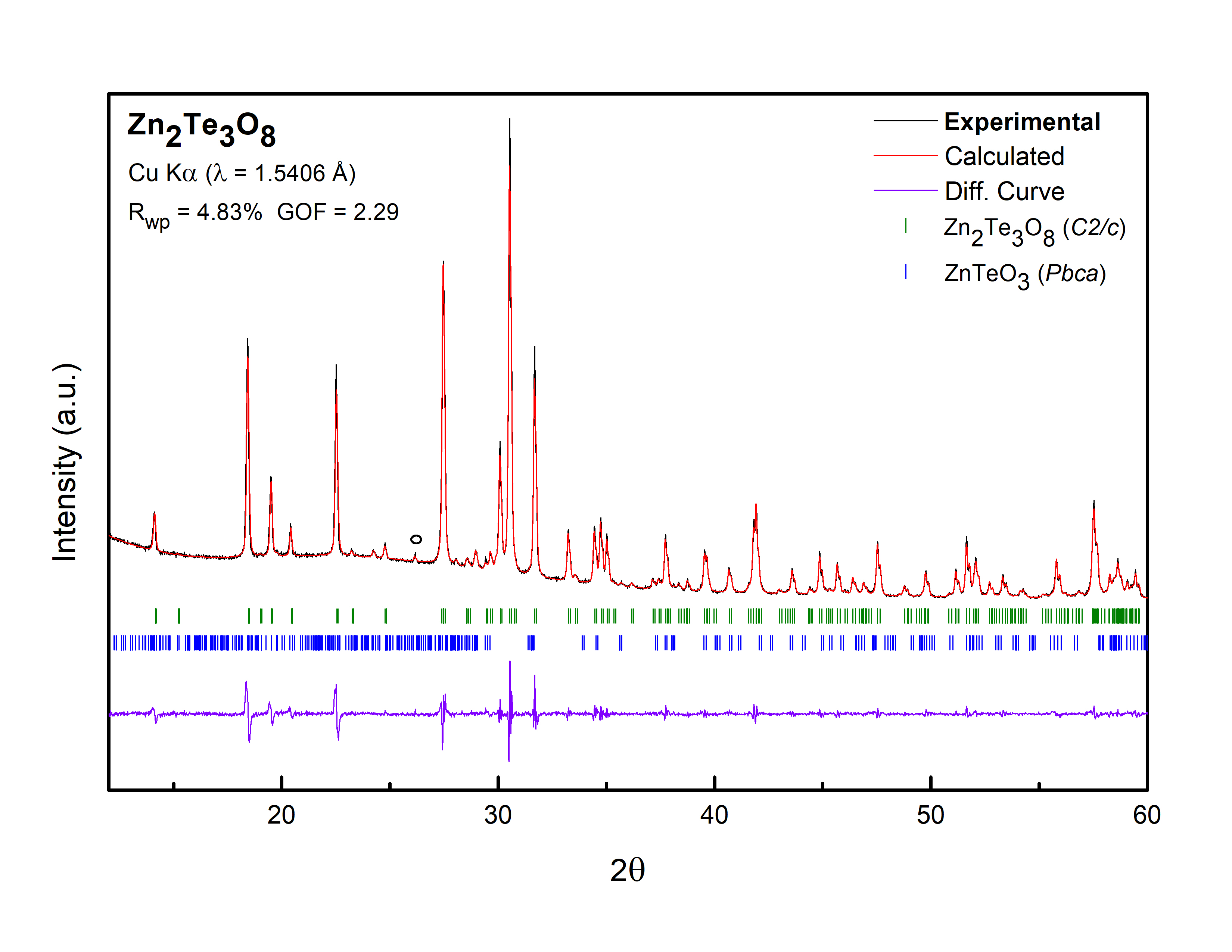}
  \caption{\textbf{} pXRD pattern of a representative \ch{Zn2Te3O8} sample synthesized in the presence of \ch{SiO2} (black). Two-phase simulated Rietveld refinement profile for \ch{Zn2Te3O8} ($C2/c$) and \ch{ZnTeO3} ($Pbca$) (red), and difference curve between the experimental and simulated patterns (purple). The black circle denotes a minor \ch{TeO2} ($P4_12_12$) impurity.} 
  \label{pXRD_Zn}
\end{figure} 
\clearpage

\begin{table}[]
\centering
\hspace*{-1.2cm}
\begin{tabular}{|c|c|c|c|c|c|c|c|}
    \hline
   \textbf{Sample} & \textbf{a(\AA)} & \textbf{b(\AA)} & \textbf{c(\AA)} & \textbf{$\beta$(°)} & \textbf{R$_{wp}$} & \textbf{R$_p$} & \textbf{GOF} \\ \hline
     AF & 12.6792(3) & 5.2081(1) & 11.6251(3) & 99.003(2) & 1.66 & 1.28 & 1.38 \\
     Li & 12.6818(6) & 5.2091(1) & 11.6257(5) & 98.987(4) & 2.39 & 1.47 & 2.13 \\
     Na (1) & 12.6760(2) & 5.2064(1) & 11.6232(2) & 99.017(2) & 1.53 & 1.20 & 1.34 \\
     Na (2) & 12.6773(4) & 5.2083(2) & 11.6248(4) & 98.995(3) & 1.97 & 1.37 & 1.80 \\
     K & 12.6780(4) & 5.2081(1) & 11.6236(3) & 99.006(2) & 2.41 & 1.55 & 2.17 \\
     Rb & 12.6727(4) & 5.2064(2) & 11.6213(4) & 99.008(3) & 1.79 & 1.29 & 1.65 \\
     Cs & 12.6763(6) & 5.2069(2) & 11.6225(5) & 99.002(4) & 1.73 & 1.26 & 1.50 \\
     Zn & 12.6807(3) & 5.2005(1) & 11.7825(3) & 99.591(2) & 4.83 & 3.28 & 2.29 \\ \hline
    \end{tabular}
    \caption{\textbf{Results of Rietveld refinement to pXRD data collected on representative \ch{Co2Te3O8} and \ch{Zn2Te3O8} samples synthesized in the presence of \ch{SiO2} and \ch{M^{+}_{2}CO3}}. All refinements were carried out in space group \emph{C2/c} (15).}
    \label{pXRD_Refinement_Params}
\end{table}

\begin{table}[]
\centering
\hspace*{-1.8cm}
\begin{tabular}{|c|c|c|c|c|c|c|c|}
    \hline
    \multicolumn{3}{|c|}{\textbf{\ch{Sample =}}} & \textbf{AF} & \textbf{Li} & \textbf{Na (1)} & \textbf{Na (2)} \\ \hline
    \multirow{4}{*}{Co1} & & x & 0.2731(7) & 0.2746(9) & 0.2710(4) & 0.2722(6) \\
    \multirow{4}{*}{(\emph{8f})} & & y & 0.3110(20) & 0.3120(30) & 0.3086(13) & 0.3161(17) \\
    & x,y,z & z & 0.1484(8) & 0.1439(10) & 0.1481(5) & 0.1477(6) \\
    & & Occ. & 0.93(1) & 0.97(1) & 0.96(1) & 0.97(1) \\
    & & B$_{eq}$ & 1.6(4) & 1.8(4) & 1.1(2) & 1.5(3) \\ \hline
    \multirow{2}{*}{Te1} & & y & 0.6613(11) & 0.6522(13) & 0.6549(6) & 0.6528(8) \\
    \multirow{2}{*}{(\emph{4e})} & $\frac{1}{2}$,y,$\frac{1}{4}$ & Occ. & 0.96(1) & 1.00(1) & 1.00(1) & 1.00(1) \\
    & & B$_{eq}$ & 1.8(3) & 1.5(3) & 1.1(2) & 1.2(2) \\ \hline
    \multirow{4}{*}{Te2} & & x & 0.3632(3) & 0.3610(4) & 0.3634(2) & 0.3617(3) \\
    \multirow{4}{*}{(\emph{8f})} & & y & 0.3086(7) & 0.3085(9) & 0.3088(4) & 0.3088(5) \\
    & x,y,z & z & 0.4440(4) & 0.4443(6) & 0.4449(3) & 0.4435(4) \\
    & & Occ. & 0.96(5) & 1.00(1) & 1.00(7) & 0.99(8) \\
    & & B$_{eq}$ & 1.9(2) & 2.4(2) & 1.2(1) & 1.7(2) \\ \hline
    \multirow{4}{*}{O1} & & x & 0.4420(30) & 0.4210(30) & 0.4172(14) & 0.4176(19) \\
    \multirow{4}{*}{(\emph{8f})} & & y & 0.4630(70) & 0.4470(70) & 0.4520(40) & 0.4610(50) \\
    & x,y,z & z & 0.1600(30) & 0.1500(40) & 0.1515(18) & 0.1540(20) \\
    & & Occ. & 1.00(4) & 0.94(3) & 0.99(2) & 0.88(2) \\
    & & B$_{eq}$ & 0.2(13) & 3.0(14) & 1.3(7) & 2.0(10) \\ \hline
    \multirow{4}{*}{O2} & & x & 0.3760(20) & 0.3670(30) & 0.3847(15) & 0.3730(20) \\
    \multirow{4}{*}{(\emph{8f})} & & y & 0.6090(6) & 0.6100(80) & 0.6150(40) & 0.6090(50) \\
    & x,y,z & z & 0.3840(30) & 0.3530(50) & 0.3560(20) & 0.3540(30) \\
    & & Occ. & 1.00(4) & 1.00(4) & 1.00(2) & 1.00(3) \\
    & & B$_{eq}$ & 0.20(12) & 3.0(14) & 2.0(8) & 1.6(9) \\ \hline
    \multirow{4}{*}{O3} & & x & 0.3000(30) & 0.2890(30) & 0.3007(14) & 0.2936(18) \\
    \multirow{4}{*}{(\emph{8f})} & & y & 0.1390(8) & 0.1410(10) & 0.1390(50) & 0.1510(60) \\
    & x,y,z & z & 0.3140(30) & 0.3280(4) & 0.3148(18) & 0.3230(20) \\
    & & Occ. & 0.95(4) & 0.87(4) & 0.93(2) & 0.93(3) \\
    & & B$_{eq}$ & 1.9(12) & 1.6(14) & 1.9(7) & 1.2(9) \\ \hline
    \multirow{4}{*}{O4} & & x & 0.2730(30) & 0.2510(40) & 0.2474(17) & 0.2540(20) \\
    \multirow{4}{*}{(\emph{8f})} & & y & 0.4820(60) & 0.4630(80) & 0.4600(40) & 0.4370(50) \\
    & x,y,z & z & 0.4810(30) & 0.4770(50) & 0.4644(18) & 0.4700(20) \\
    & & Occ. & 0.98(5) & 0.98(3) & 1.00(2) & 0.96(3) \\
    & & B$_{eq}$ & 1.1(16) & 2.9(14) & 1.7(7) & 2.0(10) \\\hline
    \end{tabular}
    \caption{\textbf{Refined atomic positions, site occupancies, and isotropic thermal parameters of representative \ch{Co2Te3O8} samples synthesized in the presence of \ch{SiO2} and \ch{M^{+}_{2}CO3}}. All refinements were carried out in space group \emph{C2/c} (15).}
    \label{pXRD_Refinement_Table1}
\end{table}

\begin{table}[]
\centering
\hspace*{-1.8cm}
\begin{tabular}{|c|c|c|c|c|c|c|c|}
    \hline
    \multicolumn{3}{|c|}{\textbf{\ch{Sample =}}} & \textbf{K} & \textbf{Rb} & \textbf{Cs} & \textbf{Zn} \\ \hline
    \multirow{4}{*}{Co1/Zn1} & & x & 0.2704(6) & 0.2715(5) & 0.2735(9) & 0.2711(5) \\
    \multirow{4}{*}{(\emph{8f})} & & y & 0.3078(18) & 0.3112(16) & 0.3180(30) & 0.2911(12) \\
    & x,y,z & z & 0.1478(7) & 0.1497(6) & 0.1433(9) & 0.1542(6) \\
    & & Occ. & 0.98(1) & 0.96(1) & 0.96(1) & 1.00(1) \\
    & & B$_{eq}$ & 1.3(3) & 1.4(3) & 1.7(4) & 0.3(2) \\ \hline
    \multirow{2}{*}{Te1} & & y & 0.6526(8) & 0.6534(7) & 0.6512(12) & 0.6381(8) \\
    \multirow{2}{*}{(\emph{4e})} & $\frac{1}{2}$,y,$\frac{1}{4}$ & Occ. & 1.00(1) & 1.00(1) & 1.00(1) & 1.00(1) \\
    & & B$_{eq}$ & 1.1(2) & 0.8(2) & 2.0(2) & 0.54(17) \\ \hline
    \multirow{4}{*}{Te2} & & x & 0.3625(3) & 0.3622(2) & 0.3609(4) & 0.3633(2) \\
    \multirow{4}{*}{(\emph{8f})} & & y & 0.3087(6) & 0.3082(5) & 0.3084(8) & 0.3041(6) \\
    & x,y,z & z & 0.4435(4) & 0.4442(3) & 0.4437(5) & 0.4452(3) \\
    & & Occ. & 0.99(9) & 0.99(8) & 0.99(1) & 1.00(1) \\
    & & B$_{eq}$ & 1.6(2) & 1.3(2) & 2.0(2) & 0.05(13) \\ \hline
    \multirow{4}{*}{O1} & & x & 0.4190(20) & 0.4169(17) & 0.4000(20) & 0.4211(19) \\
    \multirow{4}{*}{(\emph{8f})} & & y & 0.4490(60) & 0.4530(50) & 0.4620(80) & 0.4040(60) \\
    & x,y,z & z & 0.1480(30) & 0.1590(20) & 0.1470(40) & 0.1450(20) \\
    & & Occ. & 0.94(2) & 0.92(2) & 0.89(4) & 1.00(5) \\
    & & B$_{eq}$ & 1.2(10) & 2.0(9) & 1.9(16) & 1.5(10) \\ \hline
    \multirow{4}{*}{O2} & & x & 0.3770(20) & 0.3756(18) & 0.3690(30) & 0.3870(20) \\
    \multirow{4}{*}{(\emph{8f})} & & y & 0.6120(50) & 0.6010(40) & 0.6130(70) & 0.6250(50) \\
    & x,y,z & z & 0.3600(30) & 0.3600(2) & 0.3510(40) & 0.361(3) \\
    & & Occ. & 1.00(3) & 1.00(3) & 1.00(5) & 1.00(3) \\
    & & B$_{eq}$ & 0.9(10) & 1.0(8) & 1.5(14) & 0.6(9) \\ \hline
    \multirow{4}{*}{O3} & & x & 0.2954(19) & 0.2925(16) & 0.2890(30) & 0.3070(20) \\
    \multirow{4}{*}{(\emph{8f})} & & y & 0.1440(60) & 0.1480(50) & 0.1470(90) & 0.1250(50) \\
    & x,y,z & z & 0.3180(30) & 0.3200(20) & 0.3330(30) & 0.3050(20) \\
    & & Occ. & 0.94(3) & 0.97(3) & 0.92(5) & 1.00(4) \\
    & & B$_{eq}$ & 1.6(10) & 1.0(8) & 2.0(14) & 0.7(9) \\ \hline
    \multirow{4}{*}{O4} & & x & 0.2490(20) & 0.2540(20) & 0.2490(30) & 0.2370(20) \\
    \multirow{4}{*}{(\emph{8f})} & & y & 0.4580(50) & 0.4370(40) & 0.4630(80) & 0.5030(50) \\
    & x,y,z & z & 0.4730(30) & 0.4700(20) & 0.4760(40) & 0.4630(20) \\
    & & Occ. & 0.97(3) & 0.99(2) & 0.98(3) & 1.00(4) \\
    & & B$_{eq}$ & 2.0(11) & 2.0(9) & 2.0(14) & 0.6(10) \\\hline
    \end{tabular}
    \caption{\textbf{Refined atomic positions, site occupancies, and isotropic thermal parameters of representative \ch{Co2Te3O8} and \ch{Zn2Te3O8} samples synthesized in the presence of \ch{SiO2} and \ch{M^{+}_{2}CO3}}. All refinements were carried out in space group \emph{C2/c} (15).}
    \label{pXRD_Refinement_Table2}
\end{table}

\begin{table}
    \caption{{\textbf{Selected bond lengths \& angles, as well as interatomic distances of representative \ch{Co2Te3O8} samples synthesized in the presence of in the presence of \ch{SiO2} and \ch{M^{+}_{2}CO3}}.} * denotes values previously reported for \ch{Co2Te3O8}.$^{10}$}
    \label{SC_derived_params}
    \centering
    \begin{tabular}{|c|c|c|c|c|c|c|c|}
    \hline
     & \textbf{Co-O3-Co} & \textbf{Co-O4-Co} & \textbf{Co-Co (1)} & \textbf{Co-Co (2)} & \textbf{Interplanar} \\
     \textbf{Sample} & \textbf{bond angle} & \textbf{bond angle} & \textbf{bond length} & \textbf{bond length} & \textbf{Co-Co dist.}\\
     & \textbf{(degrees)} & \textbf{(degrees)} & \textbf{(\AA{})} & \textbf{(\AA{})} & \textbf{(\AA{})} \\ \hline
     Literature* & 123.10(18) & 101.63(15) & 3.6134(15) & 3.4457(18) & 5.9431(18) \\ \hline
     AF & 123.01 & 100.12 & 3.6238 & 3.4688 & 5.8677\\ \hline
     Li & 121.51 & 95.98 & 3.7074 & 3.3679 & 5.8613\\ \hline
     Na (1) & 120.71 & 92.02 & 3.6290 & 3.4614 & 5.8935\\ \hline
     Na (2) & 122.04 & 95.78 & 3.6145 & 3.4553 & 5.9155\\ \hline
     K & 121.81 & 95.90 & 3.6169 & 3.4456 & 5.9341\\ \hline
     Rb & 121.48 & 92.61 & 3.5930 & 3.4945 & 5.8945\\ \hline
     Cs & 119.64 & 95.85 & 3.7082 & 3.3656 & 5.8886\\ \hline
    \end{tabular}
\end{table} 
\clearpage

\section{EDS point and map spectra}

\begin{figure}
  \includegraphics[width=1.0\textwidth]
  {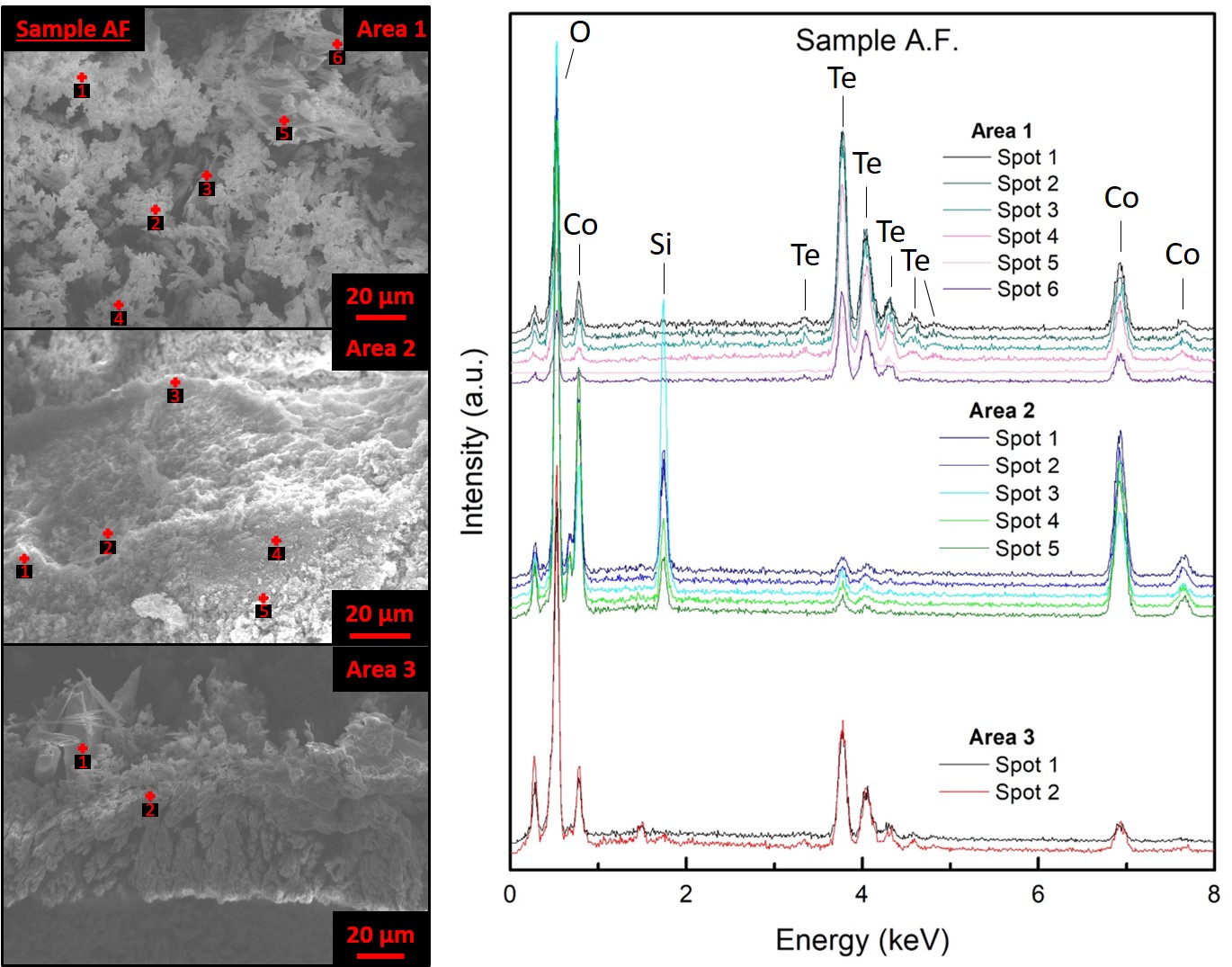}
  \caption{EDS point spectra (right) taken across three areas for Sample AF (left). The peak at 0.28 keV is present in all samples and corresponds to the characteristic carbon K$\alpha$ peak from the carbon tape used for sample mounting.} 
  \label{Sample_AF_EDS_point}
\end{figure} 

\begin{figure}
  \includegraphics[width=1.0\textwidth]
  {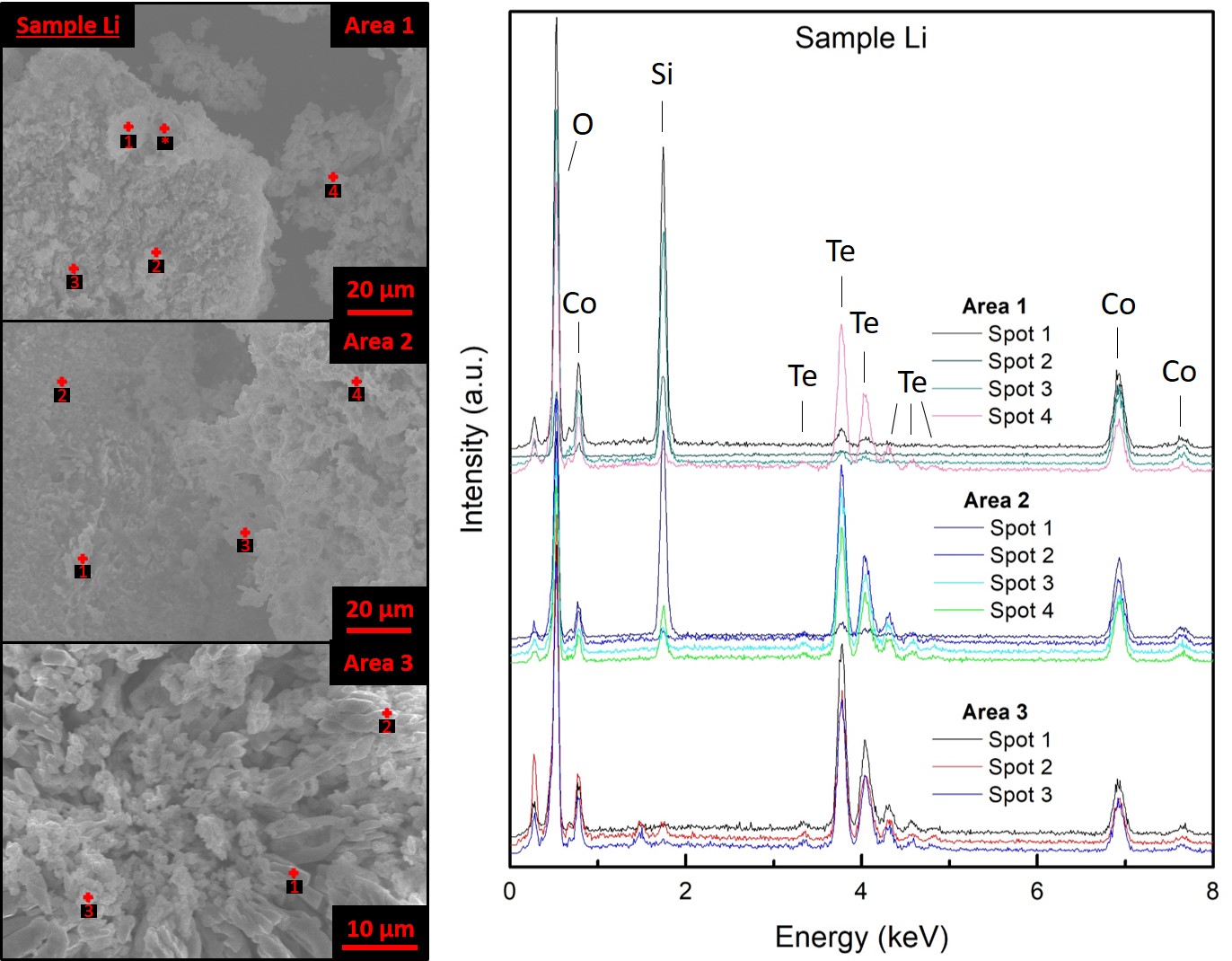}
  \caption{EDS point spectra (right) taken across three areas for Sample Li (left). The peak at 0.28 keV is present in all samples and corresponds to the characteristic carbon K$\alpha$ peak from the carbon tape used for sample mounting.} 
  \label{Sample_Li_EDS_point}
\end{figure} 

\begin{figure}
  \includegraphics[width=1.0\textwidth]
  {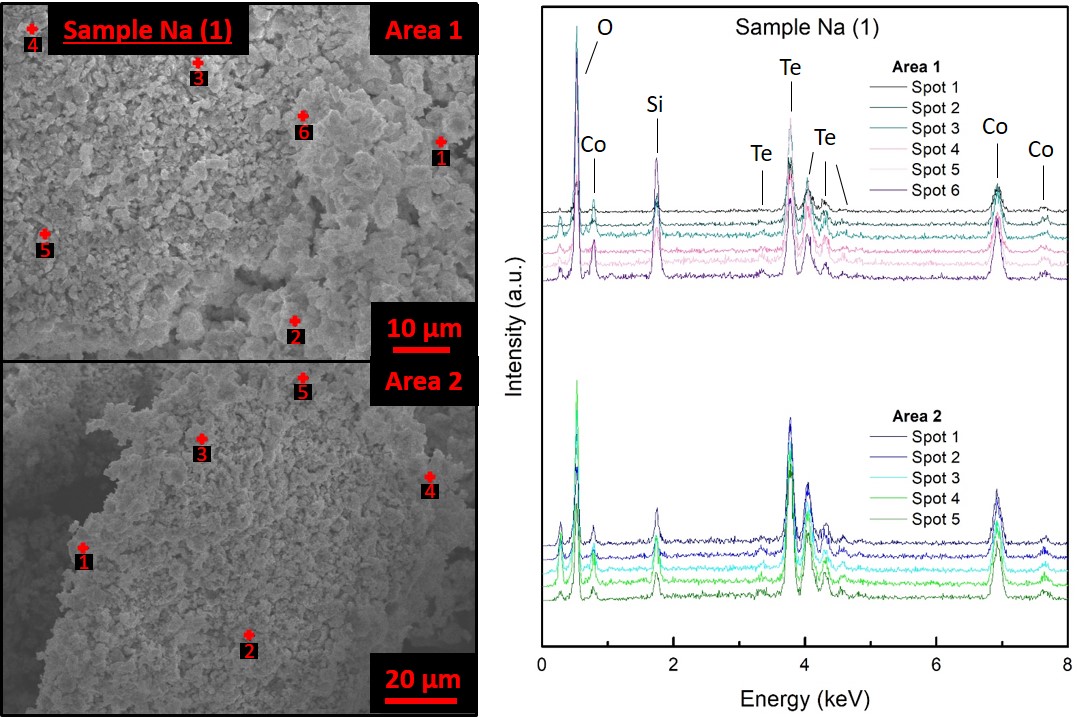}
  \caption{EDS point spectra (right) taken across two areas for Sample Na (1) (left). The peak at 0.28 keV is present in all samples and corresponds to the characteristic carbon K$\alpha$ peak from the carbon tape used for sample mounting.} 
  \label{Sample_Na1_EDS_point}
\end{figure} 

\begin{figure}
  \includegraphics[width=1.0\textwidth]
  {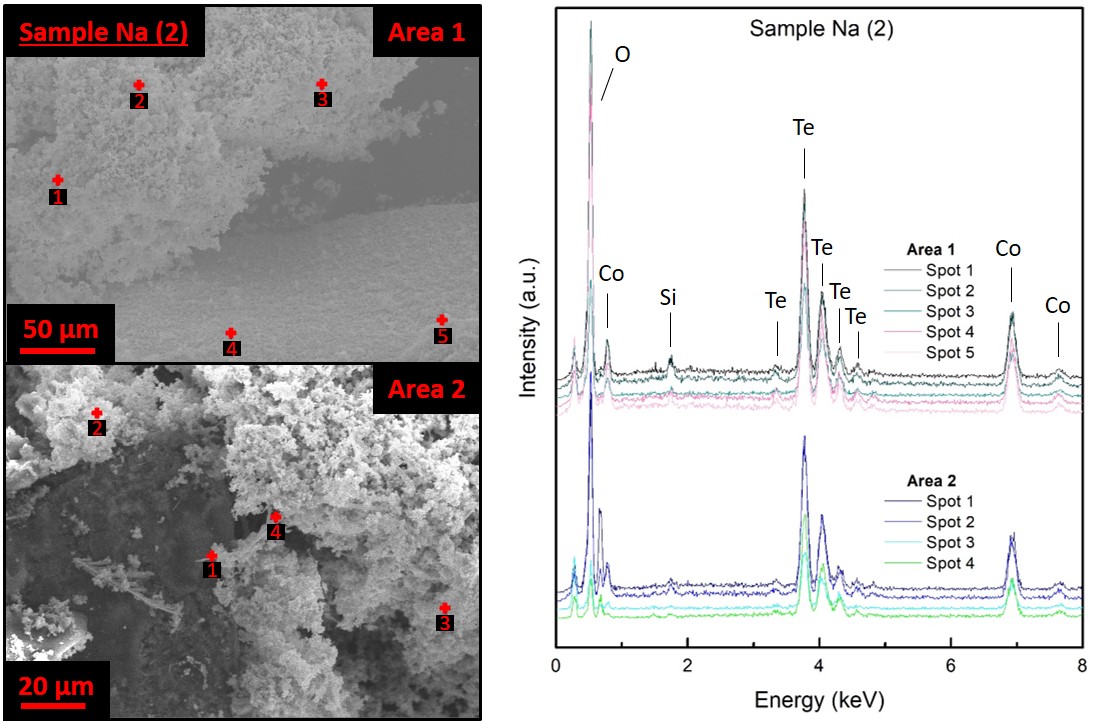}
  \caption{EDS point spectra (right) taken across two areas for Sample Na (2) (left). The peak at 0.28 keV is present in all samples and corresponds to the characteristic carbon K$\alpha$ peak from the carbon tape used for sample mounting.} 
  \label{Sample_Na2_EDS_point}
\end{figure}
\clearpage

\begin{landscape}
    \begin{table}
    \centering
    \hspace*{-0.7cm}
    \begin{tabular}{|c|c|c|c|c|c|}
    \hline
   \multicolumn{2}{|c|}{\textbf{Sample AF}} & \multicolumn{4}{c|}{\textbf{Atom \%}} \\ \hline
   \textbf{Area} & \textbf{Spot} & \textbf{Co} & \textbf{Te} & \textbf{Si} & \textbf{O} \\ \hline
   \multirow{6}{*}{1} & 1 & 19.9(2) & 26.3(5) & 0 & 53.3(1) \\
   & 2 & 18.8(2) & 29.0(5) & 0 & 52.2(1) \\
   & 3 & 19.5(2) & 30.6(5) & 0 & 49.7(1) \\
   & 4 & 24.6(3) & 39.8(5) & 0 & 35.6(1) \\
   & 5 & 41.3(5) & 53.5(9) & 0.8(1)$^*$ & 4.4(1) \\
   & 6 & 22.8(3) & 36.6(7) & 0.4(1)$^*$ & 40.3(1) \\ \hline
    \multirow{5}{*}{2} & 1 & 22.6(3) & 1.3(2) & 7.2(1) & 68.9(1) \\
   & 2 & 19.0(3) & 1.2(2) & 8.6(1) & 71.3(2) \\
   & 3 & 11.0(2) & 1.3(2) & 15.4(2) & 72.4(2) \\
   & 4 & 22.5(3) & 1.3(2) & 5.8(1) & 70.4(2) \\
   & 5 & 22.2(3) & 1.0(2) & 4.2(1) & 72.6(2) \\ \hline
    \multirow{2}{*}{3} & 1 & 6.2(2) & 17.1(6) & 0 & 76.6(2) \\
   & 2 & 8.1(2) & 16.7(5) & 0.7(1)$^*$ & 74.5(2) \\ \hline
    \hline
   \multicolumn{2}{|c|}{\textbf{Sample Li}} & \multicolumn{4}{c|}{\textbf{Atom \%}} \\ \hline
   \textbf{Area} & \textbf{Spot} & \textbf{Co} & \textbf{Te} & \textbf{Si} & \textbf{O} \\ \hline
   \multirow{4}{*}{1} & 1 & 11.2(2) & 1.0(2) & 18.1(2) & 69.5(2) \\
   & 2 & 38.0(5) & 1.0(3) & 21.0(3) & 40.0(2) \\
   & 3 & 13.0(3) & 1.0(2) & 17.6(2) & 68.5(2) \\
   & 4 & 15.1(2) & 21.1(4) & 2.4(1) & 61.4(1) \\ \hline
    \multirow{4}{*}{2} & 1 & 19.2(3) & 1.5(2) & 22.0(2) & 57.1(2) \\
   & 2 & 19.2(2) & 24.5(4) & 1.3(1) & 55.0(1) \\
   & 3 & 19.8(2) & 26.0(4) & 3.6(1) & 50.6(1) \\
   & 4 & 20.6(2) & 21.1(4) & 8.3(1) & 50.0(1) \\ \hline
    \multirow{3}{*}{3} & 1 & 15.8(2) & 25.8(4) & 0.2(1) & 58.2(1) \\
   & 2 & 12.5(2) & 19.1(4) & 2.0(1) & 64.2(1) \\
   & 3 & 15.4(2) & 21.9(4) & 0.5(1) & 62.2(1) \\\hline
   \end{tabular} 
    \begin{tabular}{|c|c|c|c|c|c|}    \hline
   \multicolumn{2}{|c|}{\textbf{Sample Na (1)}} & \multicolumn{4}{c|}{\textbf{Atom \%}} \\ \hline
   \textbf{Area} & \textbf{Spot} & \textbf{Co} & \textbf{Te} & \textbf{Si} & \textbf{O} \\ \hline
   \multirow{6}{*}{1} & 1 & 33.9(7) & 28.7(1.0) & 4.3(2) & 33.1(2) \\
   & 2 & 31.9(5) & 26.3(8) & 11.1(2) & 30.8(1) \\
   & 3 & 17.3(3) & 20.0(6) & 5.1(2) & 57.6(1) \\
   & 4 & 24.1(4) & 26.0(8) & 8.0(2) & 41.9(1) \\ 
   & 5 & 20.1(3) & 26.6(6) & 6.5(1) & 46.8(1) \\
   & 6 & 15.7(3) & 10.8(5) & 15.6(2) & 57.9(2) \\ \hline
    \multirow{5}{*}{2} & 1 & 24.3(3) & 26.7(6) & 6.8(1) & 42.3(1) \\
   & 2 & 27.1(3) & 32.6(6) & 4.0(1) & 36.3(1) \\
   & 3 & 21.0(3) & 24.2(6) & 5.9(1) & 48.9(1) \\
   & 4 & 19.5(3) & 21.5(5) & 7.2(1) & 51.7(1) \\
   & 5 & 26.9(3) & 30.5(6) & 5.8(1) & 36.8(1) \\ \hline
   
   \multicolumn{2}{|c|}{\textbf{Sample Na (2)}} & \multicolumn{4}{c|}{\textbf{Atom \%}} \\ \hline
   \textbf{Area} & \textbf{Spot} & \textbf{Co} & \textbf{Te} & \textbf{Si} & \textbf{O} \\ \hline
   \multirow{5}{*}{1} & 1 & 17.4(2) & 23.3(4) & 1.5(1) & 57.9(1) \\
   & 2 & 16.2(2) & 19.3(4) & 2.4(1) & 62.2(1) \\
   & 3 & 22.8(3) & 28.6(6) & 1.2(1) & 47.4(1) \\
   & 4 & 17.5(2) & 22.7(4) & 0.5(1) & 59.3(1) \\ 
   & 5 & 15.5(2) & 20.4(4) & 1.4(1) & 62.8(1) \\ \hline
    \multirow{5}{*}{2} & 1 & 20.9(3) & 24.0(5) & 0.8(1) & 54.3(1) \\
   & 2 & 19.0(2) & 24.8(5) & 1.4(1) & 54.9(1) \\
   & 3 & 28.0(4) & 33.0(8) & 1.1(1) & 37.9(1) \\
   & 4 & 31.6(3) & 41.4(6) & 1.2(1) & 25.9(1) \\ \hline
\end{tabular}
    \caption{Calculated elemental ratios and statistical errors for Samples AF, Li, Na (1), and Na (2) estimated from energy dispersive x-ray (EDS) point spectra. $^*$ denotes a calculated Si atomic percentage for a given point spectrum where the characteristic Si peak is absent.}
    \label{All_point_spectra_1}
    \end{table}
\clearpage
\end{landscape}

\begin{figure}
  \includegraphics[width=1.0\textwidth]
  {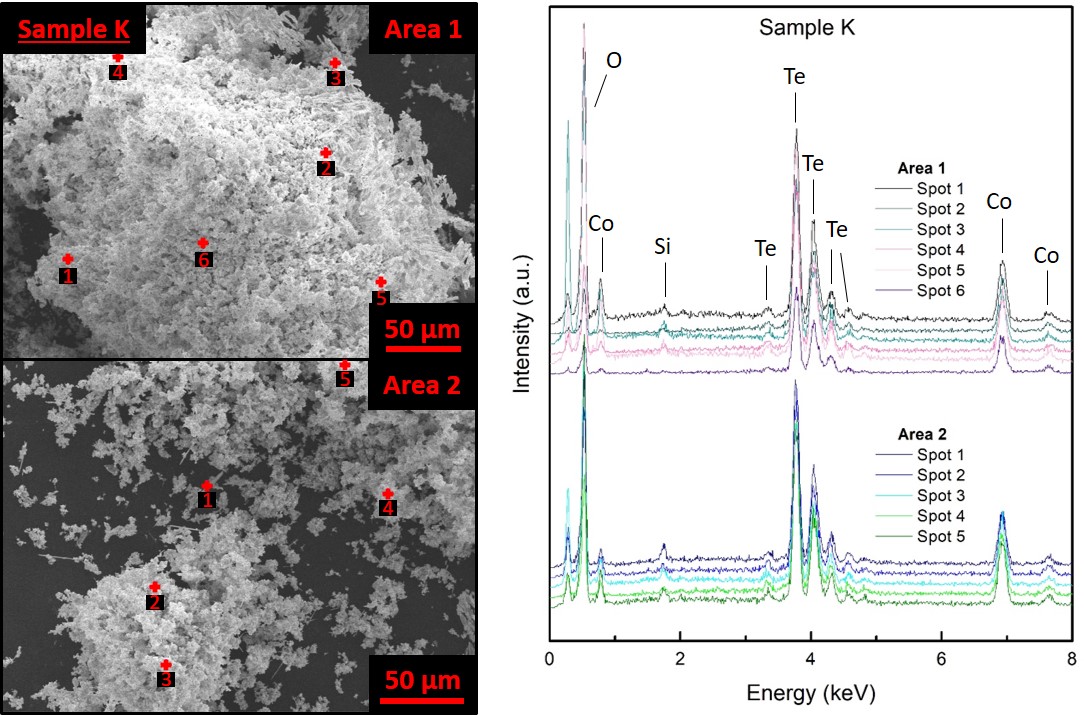}
  \caption{EDS point spectra (right) taken across two areas for Sample K (left). The peak at 0.28 keV is present in all samples and corresponds to the characteristic carbon K$\alpha$ peak from the carbon tape used for sample mounting.} 
  \label{Sample_K_EDS_point}
\end{figure} 

\begin{figure}
  \includegraphics[width=1.0\textwidth]
  {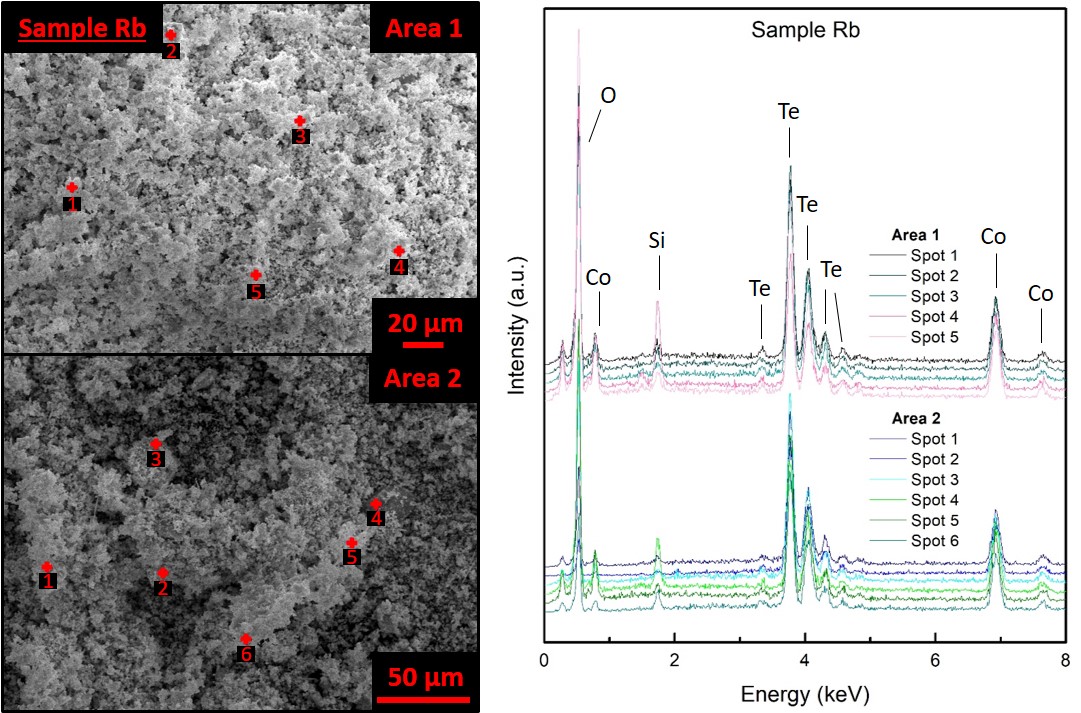}
  \caption{EDS point spectra (right) taken across two areas for Sample Rb (left). The peak at 0.28 keV is present in all samples and corresponds to the characteristic carbon K$\alpha$ peak from the carbon tape used for sample mounting.} 
  \label{Sample_Rb_EDS_point}
\end{figure} 

\begin{figure}
  \includegraphics[width=1.0\textwidth]
  {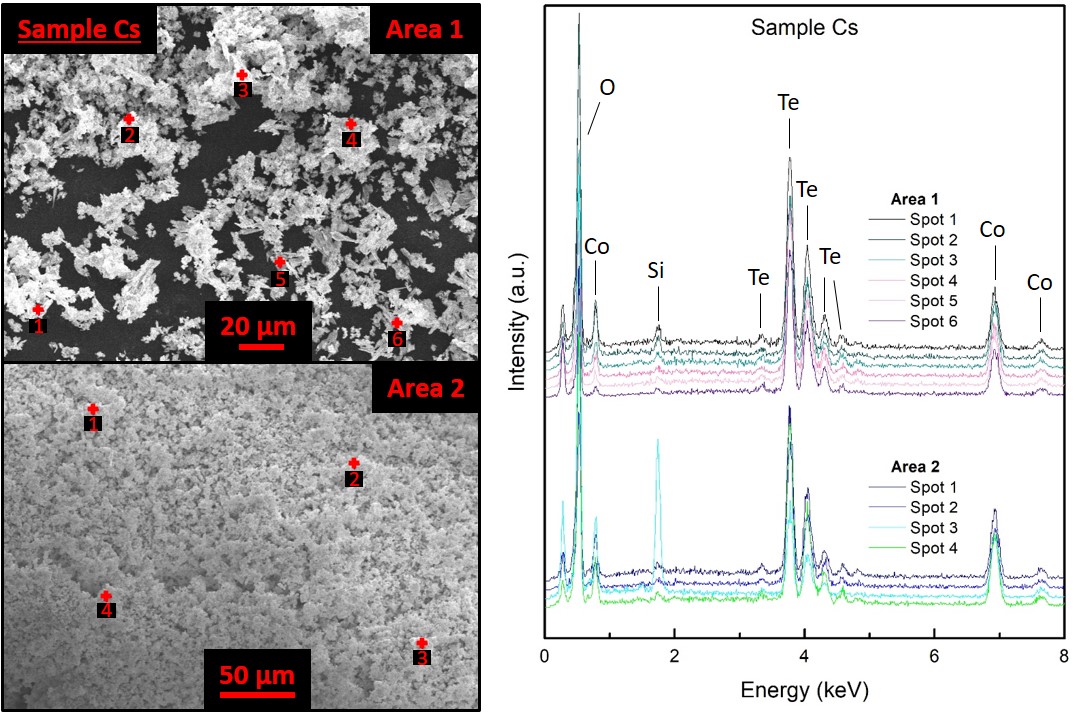}
  \caption{EDS point spectra (right) taken across two areas for Sample Cs (left). The peak at 0.28 keV is present in all samples and corresponds to the characteristic carbon K$\alpha$ peak from the carbon tape used for sample mounting.} 
  \label{Sample_Cs_EDS_point}
\end{figure} 
\clearpage

\begin{table}[]
\centering
\hspace*{-0.7cm}
\begin{tabular}{|c|c|c|c|c|c|} \hline
   \multicolumn{2}{|c|}{\textbf{Sample K}} & \multicolumn{4}{c|}{\textbf{Atom \%}} \\ \hline
   \textbf{Area} & \textbf{Spot} & \textbf{Co} & \textbf{Te} & \textbf{Si} & \textbf{O} \\ \hline
   \multirow{6}{*}{1} & 1 & 17.2(2) & 23.7(4) & 0.8(1) & 58.3(1) \\
   & 2 & 33.5(3) & 45.3(6) & 0.8(1) & 20.4(1) \\
   & 3 & 15.0(2) & 22.4(5) & 1.8(1) & 60.8(1) \\
   & 4 & 26.1(3) & 39.5(5) & 1.0(1) & 33.4(1) \\ 
   & 5 & 17.1(2) & 24.1(4) & 1.5(1) & 57.4(1) \\ 
   & 6 & 35.5(4) & 41.2(7) & 1.0(1) & 22.3(1) \\ \hline
    \multirow{5}{*}{2} & 1 & 18.3(2) & 31.0(5) & 2.5(1) & 48.2(1) \\
   & 2 & 22.8(2) & 30.5(5) & 0.9(1) & 45.8(1) \\
   & 3 & 19.4(2) & 23.0(5) & 1.5(1) & 56.1(1) \\
   & 4 & 26.5(3) & 34.6(5) & 1.1(1) & 37.8(1) \\
   & 5 & 19.3(2) & 24.3(5) & 1.4(1) & 55.0(1) \\ \hline
   \multicolumn{2}{|c|}{\textbf{Sample Rb}} & \multicolumn{4}{c|}{\textbf{Atom \%}} \\ \hline
   \textbf{Area} & \textbf{Spot} & \textbf{Co} & \textbf{Te} & \textbf{Si} & \textbf{O} \\ \hline
   \multirow{5}{*}{1} & 1 & 19.2(2) & 24.4(4) & 1.7(1) & 54.7(1) \\
   & 2 & 18.9(2) & 25.3(4) & 1.9(1) & 54.0(1) \\
   & 3 & 19.8(2) & 26.1(4) & 1.3(1) & 52.7(1) \\
   & 4 & 18.0(2) & 15.1(4) & 9.3(1) & 57.6(1) \\ 
   & 5 & 16.1(2) & 21.4(4) & 1.5(1) & 60.9(1) \\ \hline
    \multirow{6}{*}{2} & 1 & 27.3(3) & 35.5(5) & 1.4(1) & 35.8(1) \\
   & 2 & 36.1(3) & 44.7(6) & 1.0(1) & 18.3(1) \\
   & 3 & 21.0(2) & 27.5(4) & 1.0(1) & 50.5(1) \\
   & 4 & 18.5(2) & 17.2(4) & 6.4(1) & 57.9(1) \\
   & 5 & 18.7(2) & 23.0(4) & 1.3(1) & 57.0(1) \\
   & 6 & 28.2(3) & 31.3(5) & 4.3(1) & 36.3(1) \\ \hline
   \multicolumn{2}{|c|}{\textbf{Sample Cs}} & \multicolumn{4}{c|}{\textbf{Atom \%}} \\ \hline
   \textbf{Area} & \textbf{Spot} & \textbf{Co} & \textbf{Te} & \textbf{Si} & \textbf{O} \\ \hline
   \multirow{6}{*}{1} & 1 & 15.8(2) & 24.3(4) & 1.9(1) & 58.7(1) \\
   & 2 & 15.2(2) & 19.5(4) & 2.1(1) & 63.3(1) \\
   & 3 & 19.8(2) & 26.3(5) & 1.3(1) & 52.7(1) \\
   & 4 & 19.8(2) & 28.8(5) & 0.9(1) & 50.5(1) \\ 
   & 5 & 21.3(2) & 33.7(5) & 1.1(1) & 44.0(1) \\
   & 6 & 23.3(3) & 34.8(5) & 1.3(1) & 40.6(1) \\ \hline
    \multirow{4}{*}{2} & 1 & 18.2(2) & 21.4(4) & 1.0(1) & 59.4(1) \\
   & 2 & 22.6(3) & 26.4(5) & 1.0(1) & 50.1(1) \\
   & 3 & 12.0(2) & 7.8(3) & 12.8(2) & 67.5(1) \\
   & 4 & 19.9(2) & 23.9(4) & 0.9(1) & 55.3(1) \\ \hline
    \end{tabular} \\
    \caption{Calculated elemental ratios and statistical errors for Samples K, Rb, and Cs estimated from energy dispersive x-ray (EDS) point spectra.}
    \label{All_point_spectra_2}
\end{table}
\clearpage

\begin{figure}
  \includegraphics[width=1.0\textwidth]
  {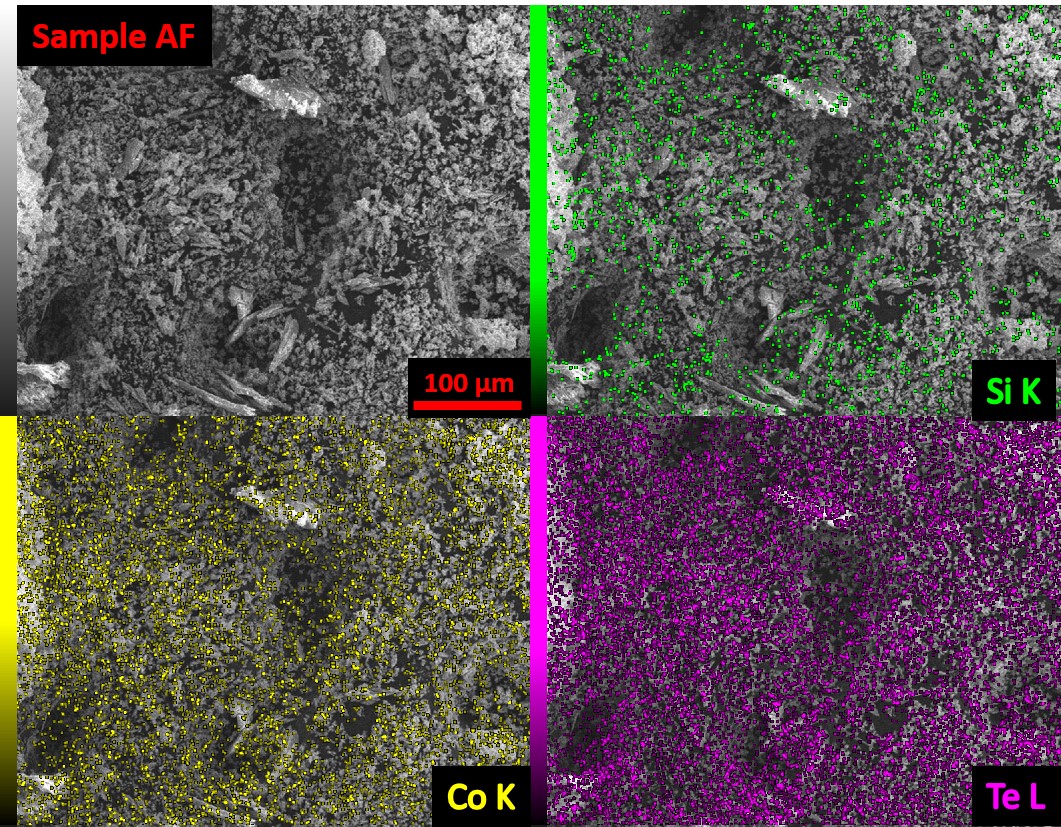}
  \caption{SED image and corresponding EDS maps (Co - yellow, Te - pink, Si - green) of a representative region of Sample AF.} 
  \label{Sample_AF_EDS_map}
\end{figure} 
\clearpage

\begin{figure}
  \includegraphics[width=1.0\textwidth]
  {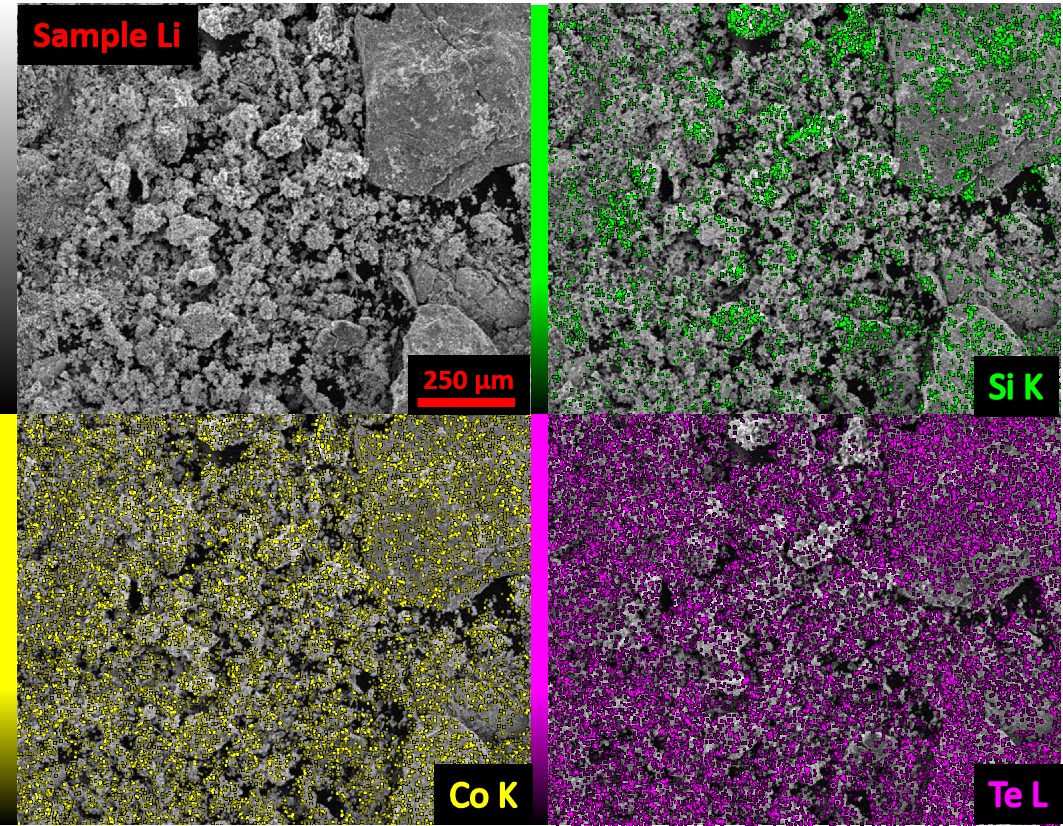}
  \caption{SED image and corresponding EDS maps (Co - yellow, Te - pink, Si - green) of a representative region of Sample Li.} 
  \label{Sample_Li_EDS_map}
\end{figure} 
\clearpage

\begin{figure}
  \includegraphics[width=1.0\textwidth]
  {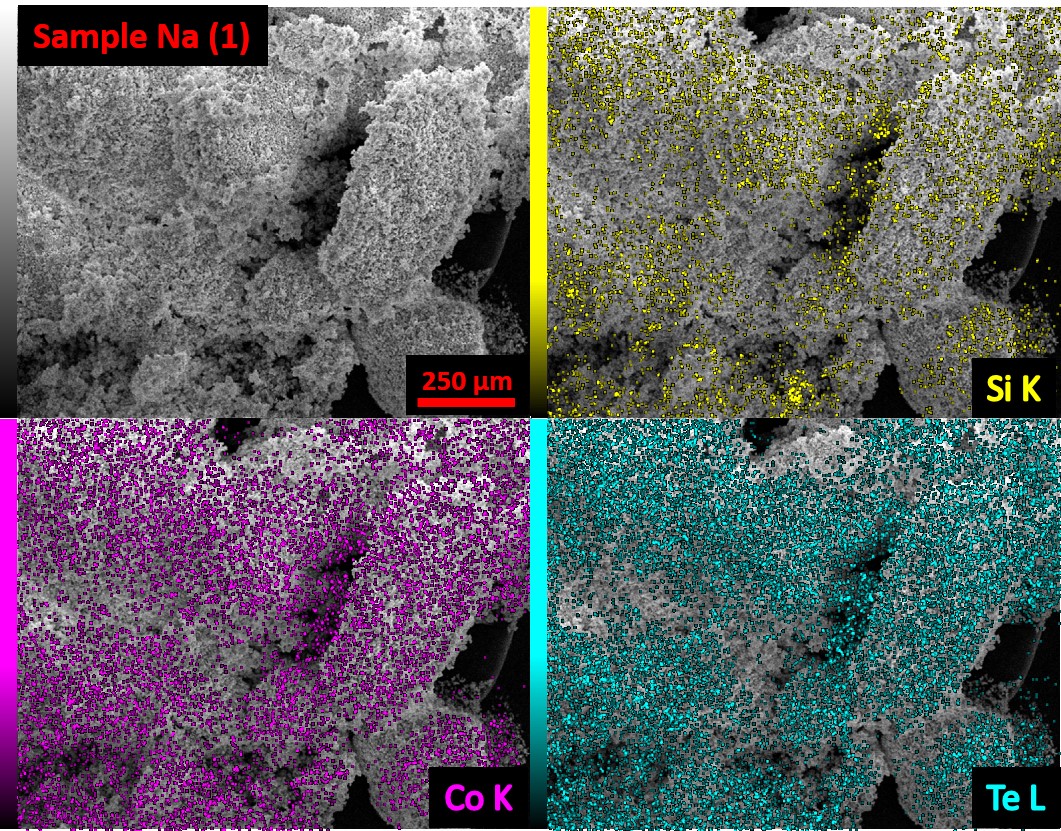}
  \caption{SED image and corresponding EDS maps (Co - pink, Te - blue, Si - green) of a representative region of Sample Na (1).} 
  \label{Sample_Na1_EDS_map}
\end{figure} 
\clearpage

\begin{figure}
  \includegraphics[width=1.0\textwidth]
  {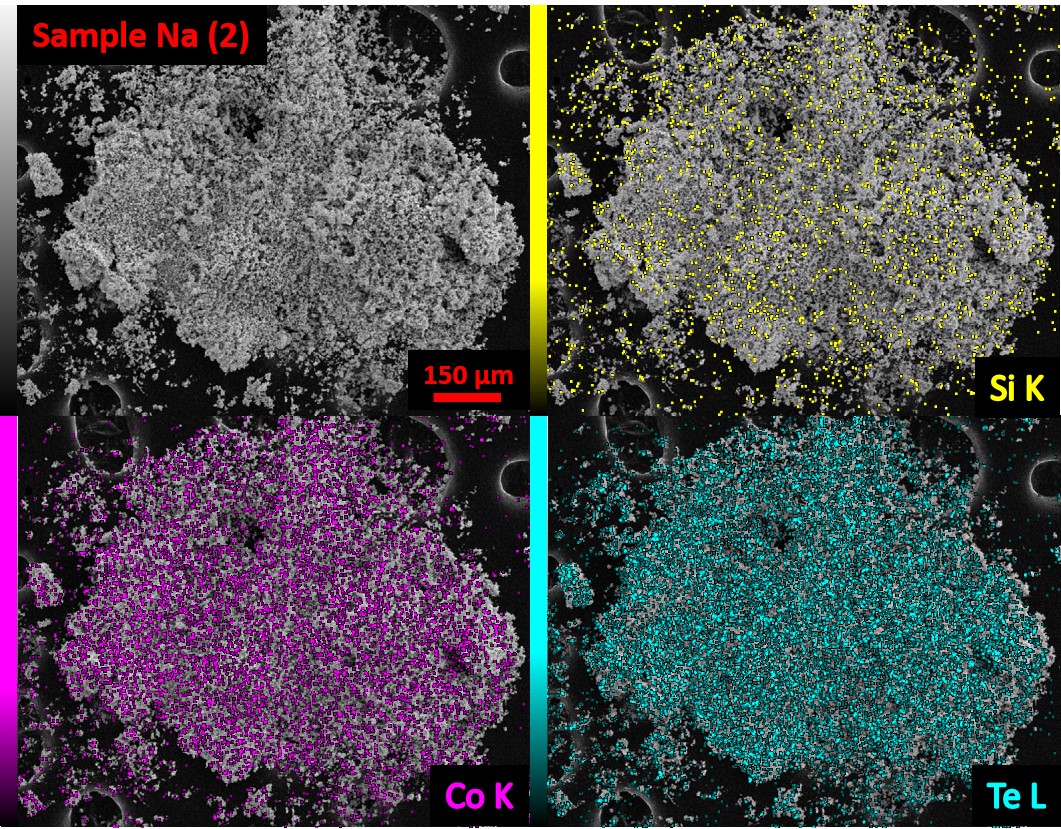}
  \caption{SED image and corresponding EDS maps (Co - pink, Te - blue, Si - yellow) of a representative region of Sample Na (2).} 
  \label{Sample_Na2_EDS_map}
\end{figure} 
\clearpage

\begin{figure}
  \includegraphics[width=1.0\textwidth]
  {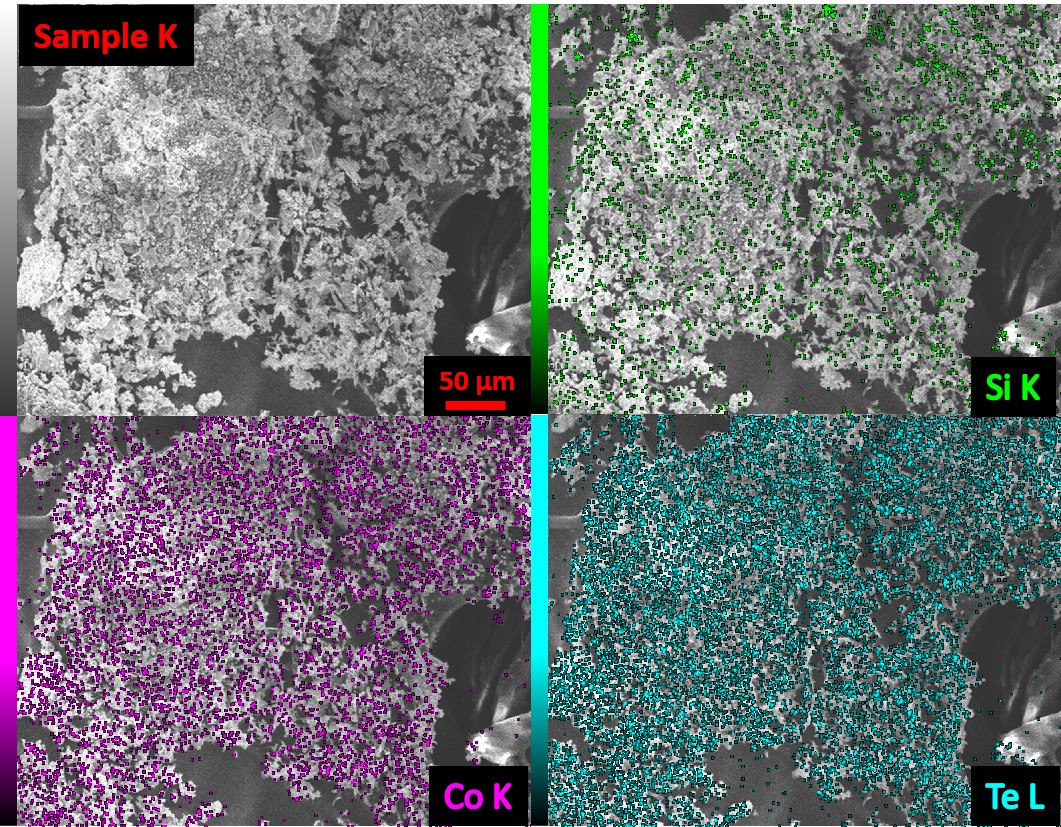}
  \caption{SED image and corresponding EDS maps (Co - pink, Te - blue, Si - green) of a representative region of Sample K.} 
  \label{Sample_K_EDS_map}
\end{figure} 
\clearpage

\begin{figure}
  \includegraphics[width=1.0\textwidth]
  {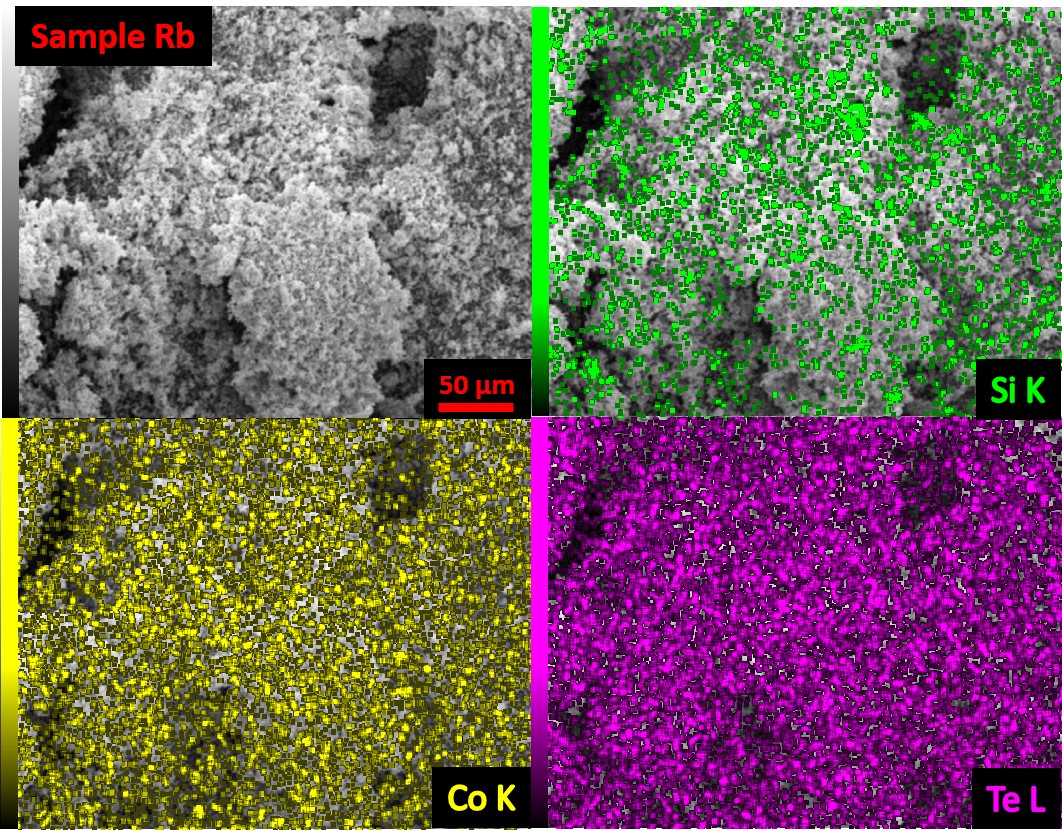}
  \caption{SED image and corresponding EDS maps (Co - yellow, Te - blue, Si - green) of a representative region of Sample Rb.} 
  \label{Sample_Rb_EDS_map_main}
\end{figure} 
\clearpage

\begin{figure}
  \includegraphics[width=1.0\textwidth]
  {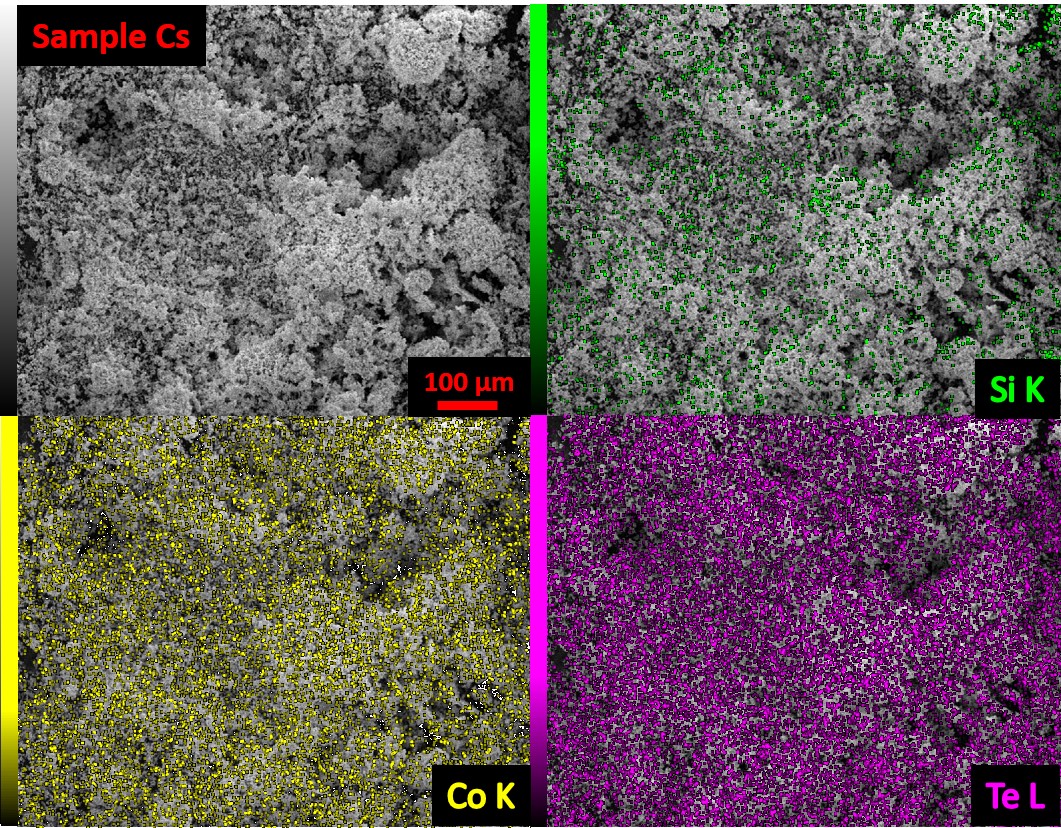}
  \caption{SED image and corresponding EDS maps (Co - yellow, Te - pink, Si - green) of a representative region of Sample Cs.} 
  \label{Sample_Cs_EDS_map}
\end{figure} 

\begin{table}[]
\centering
\hspace*{-0.7cm}
\begin{tabular}{|c|c|c|c|c|c|}
    \hline
   \multicolumn{2}{|c|}{} & \multicolumn{4}{c|}{\textbf{Atom \%}} \\ \hline
   \textbf{Sample} & \textbf{Map} & \textbf{Co} & \textbf{Te} & \textbf{Si} & \textbf{O} \\ \hline
   \multirow{2}{*}{AF} & 1 & 18.3(1) & 26.8(3) & 0.3(1) & 54.6(1) \\
   & 2 & 21.2(1) & 29.7(3) & 2.0(1) & 47.1(1) \\ \hline
   Li & 1 & 18.7(1) & 21.6(3) & 6.5(1) & 53.2(1) \\ \hline
   Na (1) & 1 & 20.7(1) & 21.6(3) & 7.2(1) & 49.3(1) \\ \hline
   \multirow{2}{*}{Na (2)} & 1 & 19.7(2) & 24.2(3) & 0.6(1) & 55.4(1) \\
   & 2 & 19.9(1) & 27.1(2) & 1.5(1) & 51.2(1) \\ \hline
   K & 1 & 17.9(2) & 23.9(3) & 3.0(1) & 55.1(1) \\ \hline
   Rb & 1 & 21.0(1) & 26.1(3) & 2.6(1) & 50.3(1) \\ \hline
   Cs & 1 & 21.9(1) & 25.9(3) & 1.9(1) & 50.3(1) \\ \hline
    \end{tabular}
    \caption{\textbf{Calculated average elemental ratios and statistical errors for all samples, estimated from energy dispersive x-ray (EDS) map spectra.} }
    \label{CTO_map_spectra}
\end{table}
\clearpage

\begin{figure}
  \includegraphics[width=1.0\textwidth]
  {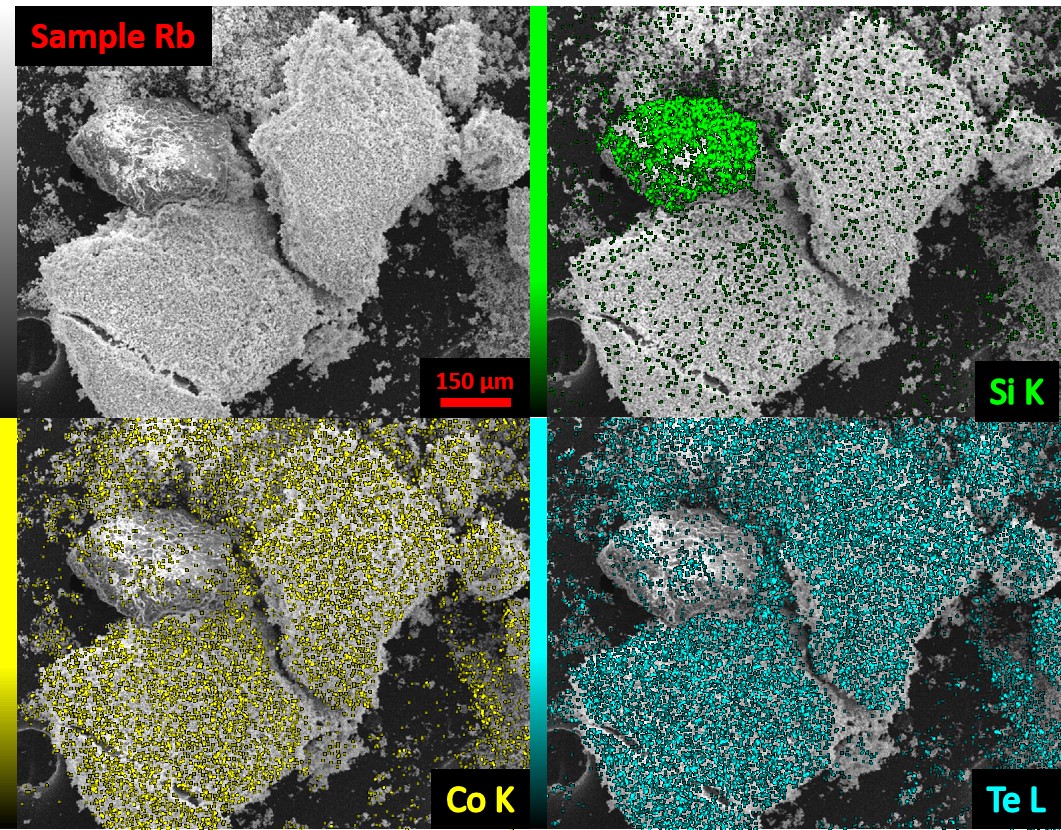}
  \caption{SED image and corresponding EDS maps (Co - yellow, Te - blue, Si - green) of a region of Sample Rb, showing the presence of a chunk of \ch{SiO2} amongst the Si-containing \ch{Co2Te3O8} phase.} 
  \label{Sample_Rb_EDS_map_SiO2}
\end{figure} 
\clearpage

\section{TGA-DTA data}

\begin{figure}
  \includegraphics[width=1.0\textwidth]
  {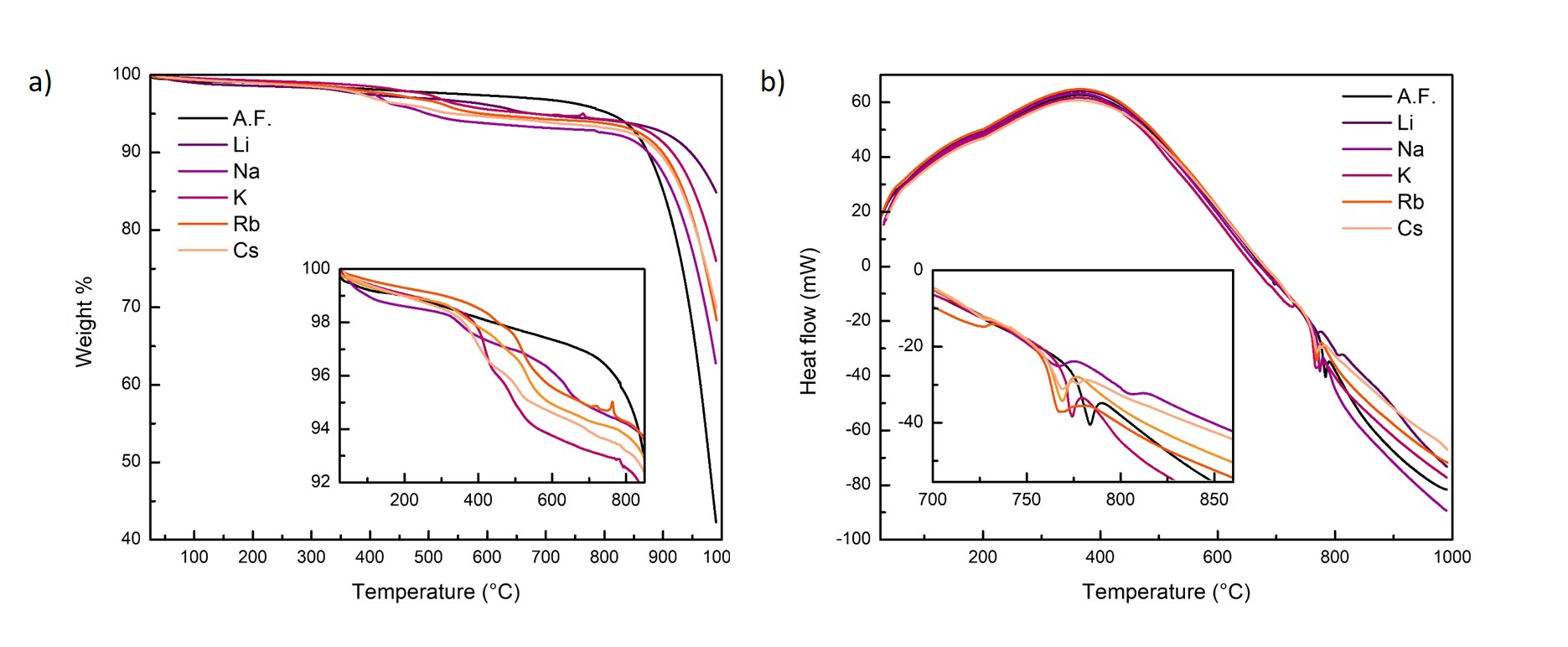}
  \caption{\textbf{Evolution of \ch{Co2Te3O8} thermal dehydration and decomposition as a function of secondary mineralizer.} a) Percent change in sample mass and b) heat flow as a function of temperature for unheated \ch{Co2Te3O8} samples.} 
  \label{TGA-DSC_comb}
\end{figure} 

\begin{figure}
  \includegraphics[width=1.0\textwidth]
  {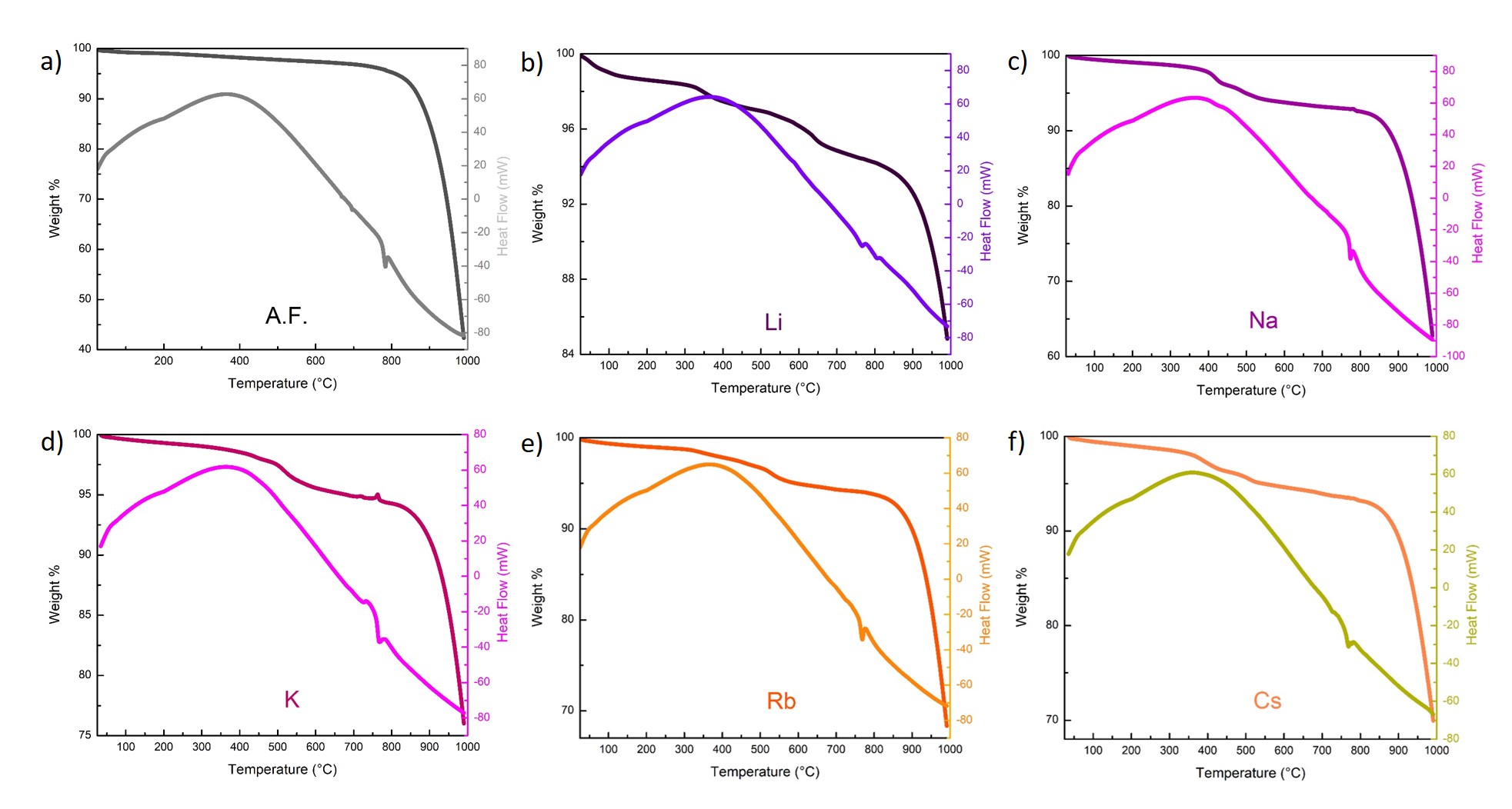}
  \caption{\textbf{} Percent change in sample mass (left) and heat flow (right) of representative \ch{Co2Te3O8} samples synthesized in the presence of \ch{SiO2} and \ch{M^{+}_{2}CO3} [M = a) none, b) Li, c) Na, d) K, e) Rb, f) Cs] as a function of temperature.} 
  \label{TGA_DSC_indiv}
\end{figure} 
\clearpage

\section{Infrared and Raman spectra}

\begin{figure}
  \includegraphics[width=1.0\textwidth]
  {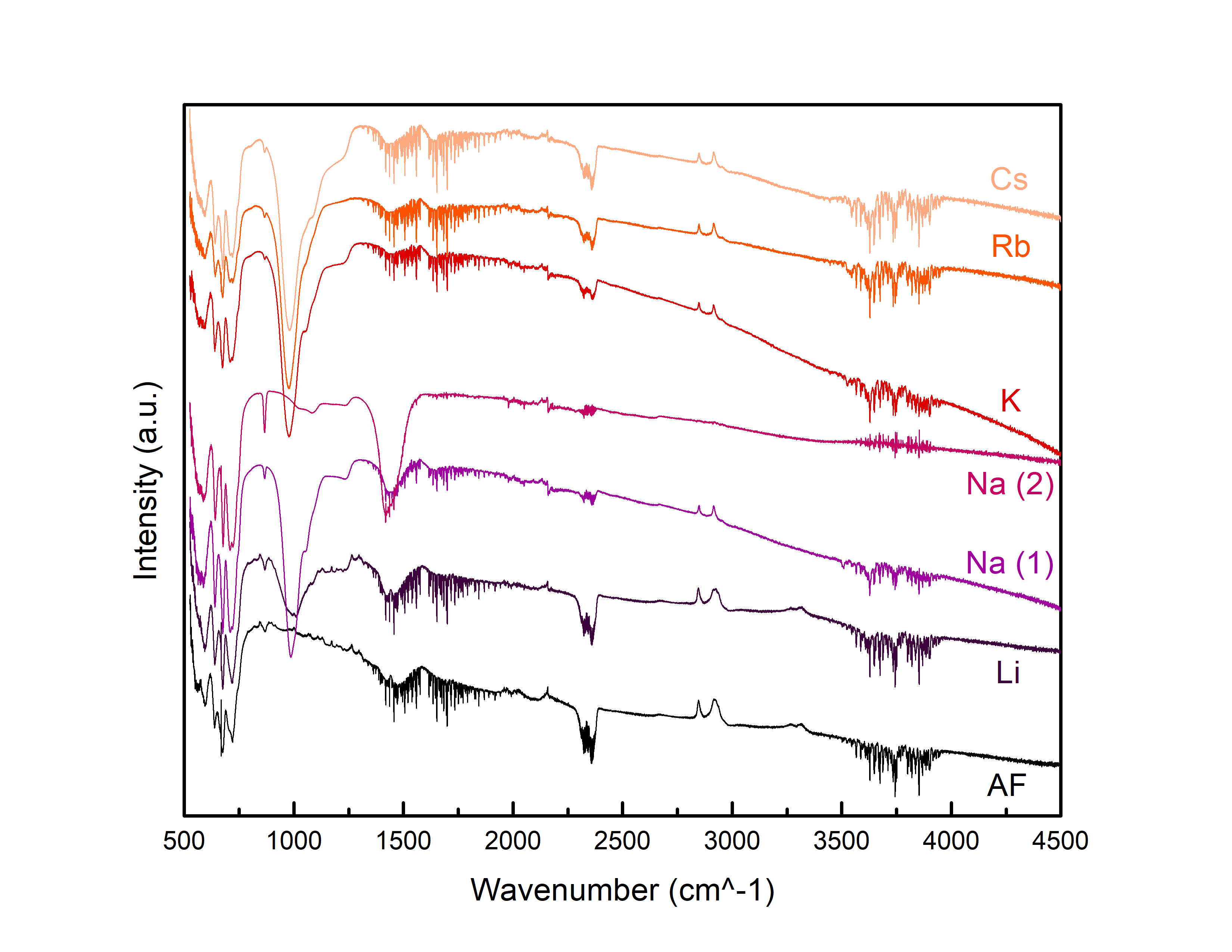}
  \caption{\textbf{} Representative high-frequency infrared spectra of (from bottom to top) Samples AF, Li, Na, K, Rb, and Cs after heating at 200°C.} 
  \label{IR_high_freq}
\end{figure} 

\begin{figure}
  \includegraphics[width=1.0\textwidth]
  {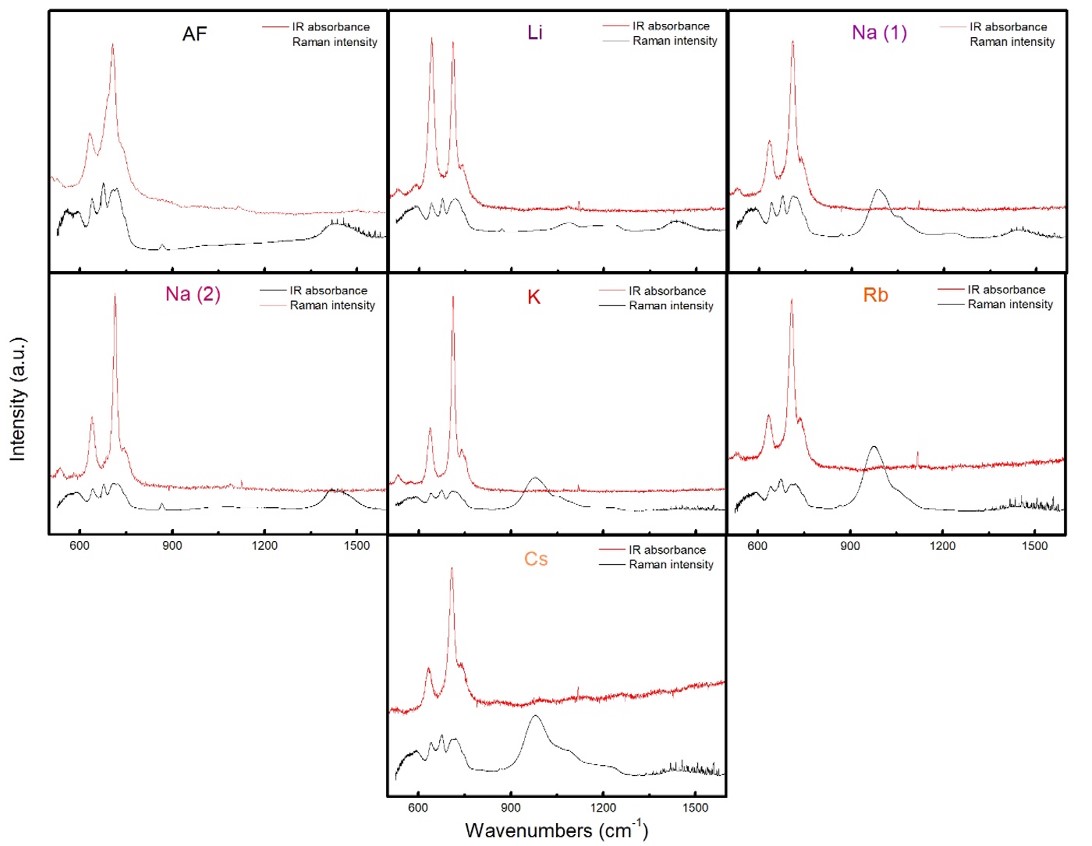}
  \caption{\textbf{} Representative infrared (lighter) and Raman scattering (darker) spectra of Samples AF (top left), Li (top center), Na (top right), K (bottom left), Rb (bottom center), and Cs (bottom right) after heating at 200°C.} 
  \label{IR_Raman_indiv}
\end{figure} 

\begin{figure}
  \includegraphics[width=1.0\textwidth]
  {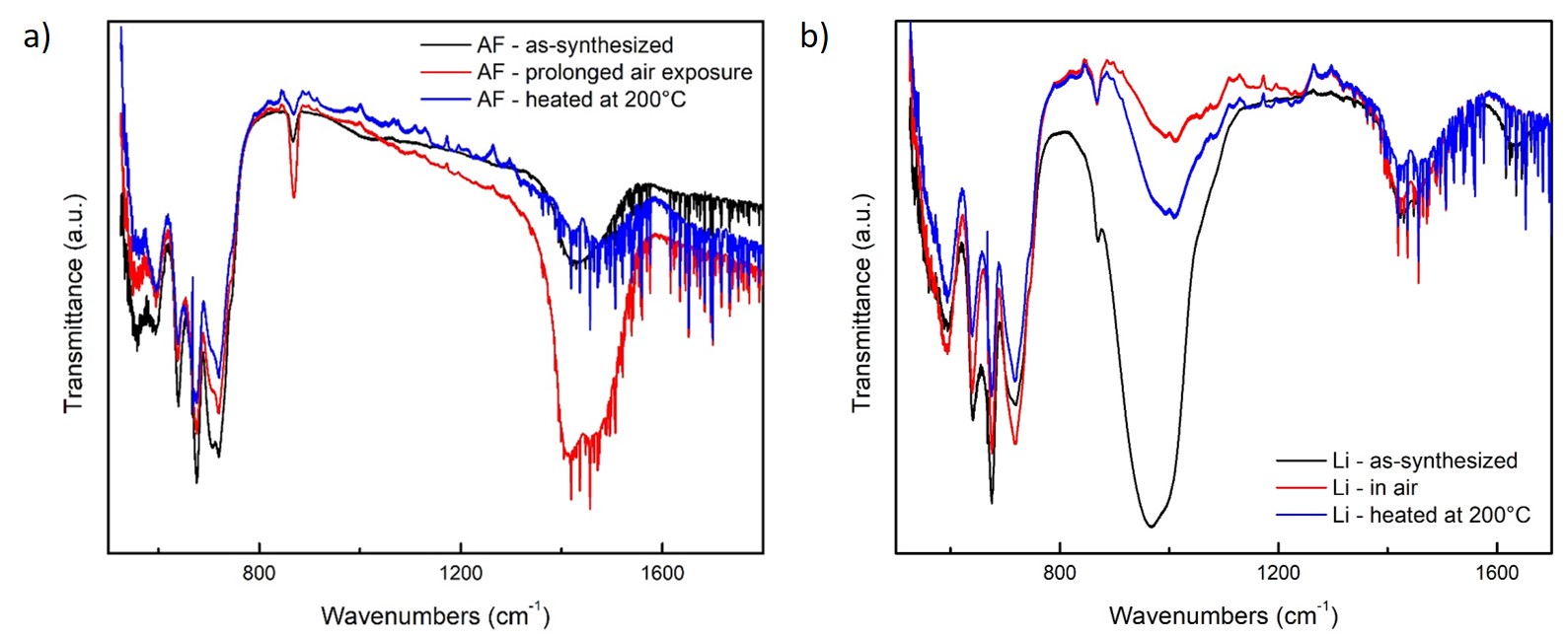}
  \caption{\textbf{} Representative infrared spectra of a) Sample AF and b) Sample Li before (black) and after (blue) heating to 200°C, as well as after prolonged air exposure (red).} 
  \label{CTO_LI_IR_comp}
\end{figure} 
\clearpage

\section{Magnetic characterization}

\begin{figure}
  \includegraphics[width=1.0\textwidth]
  {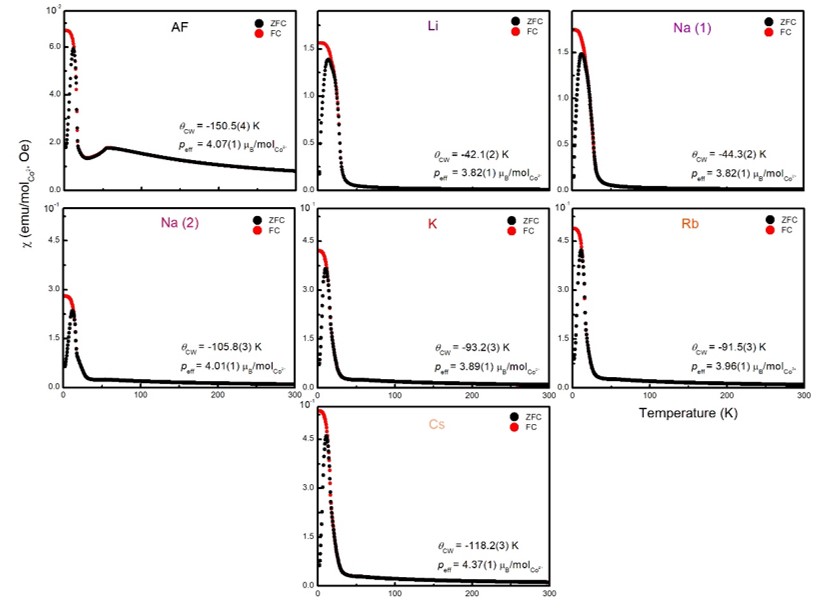}
  \caption{\textbf{} Magnetic susceptibility as a function of temperature of Samples AF (top left), Li (top center), Na (top right), K (bottom left), Rb (bottom center), and Cs (bottom right) after heating at 200°C.} 
  \label{All_dried_MvT_indiv}
\end{figure} 

\begin{table}
    \caption{\textbf{Curie-Weiss fitting parameters obtained from analysis of \ch{Co2Te3O8} powder magnetization measurements from \emph{T}~=~70-300~K.}}
    \label{CW_fits_hydration}
    \begin{tabular}{|c|c|c|c|c|}
    \hline
    \textbf{Sample} & \textbf{Drying Temp.} & \textbf{$\theta_{\text{CW}}$} & \textbf{$p_{\text{eff}}$} & \textbf{$\chi_0$} \\ 
    & (°C) & (K) & ($\mu_\text{B}$/\ch{Co^{2+}}) & \\\hline
    Na (1) & As-synthesized & -32.3(3) & 3.00(1) & -3.1(3.1)~x~10$^{-5}$ \\\hline 
    Na (1) & 200 & -44.3(2) & 3.99(1) & -3.1(3.1)~x~10$^{-5}$ \\\hline
    Na (1) & 600 & -39.3(3) & 3.48(1) & -3.1(3.1)~x~10$^{-5}$  \\\hline
    K & As-synthesized & -17.2(2) & 3.48(1) & -3.1(3.1)~x~10$^{-5}$ \\\hline
    K & 200 & -93.2(3) & 3.89(1) & -3.1(3.1)~x~10$^{-5}$ \\\hline
    Rb & As-synthesized & -9.9(2) & 3.77(1) & -3.1(3.1)~x~10$^{-5}$ \\\hline
    Rb & 200 & -91.5(3) & 3.96(1) & -3.1(3.1)~x~10$^{-5}$ \\\hline
    \end{tabular}
\end{table}

\begin{figure}
  \includegraphics[width=1.0\textwidth]
  {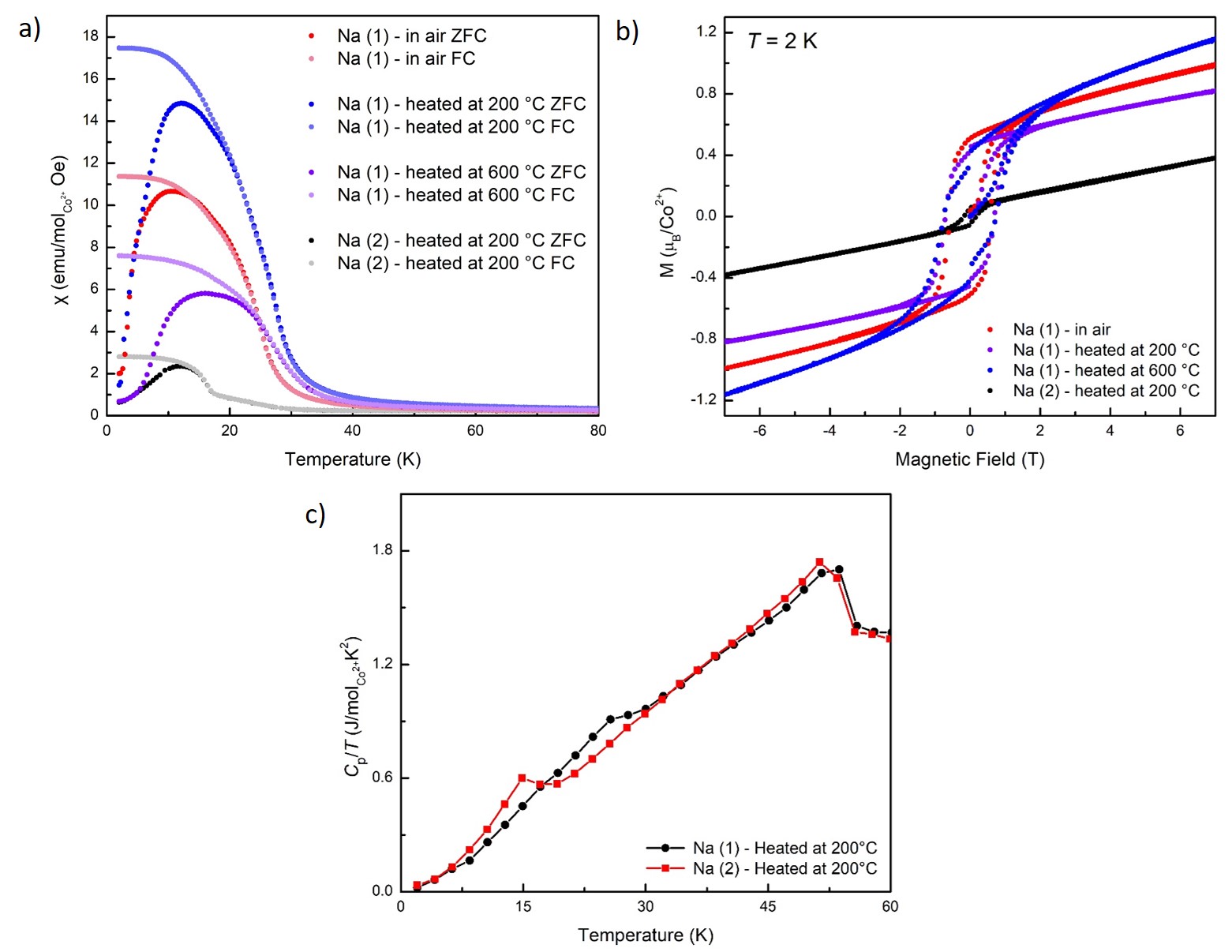}
  \caption{\textbf{} a) Zero-field-cooled (darker) and field-cooled (lighter) magnetic susceptibility as a function of temperature and b) magnetization as a function of applied field of Sample Na (1) after prolonged air exposure (red), heating at 200°C, and heating at 600°C, as well as of Sample Na (2) after heating at 200°C. c) Heat capacity divided by temperature as a function of temperature of Sample Na (1) (black) and Sample Na (2) (red) after heating at 200°C.} 
  \label{NCTeCO_all_comb}
\end{figure} 
\clearpage